\documentclass[%
reprint,
superscriptaddress,
prb,
aps
]{revtex4-2}

\usepackage{amssymb}
\usepackage{amsmath}
\usepackage{bbold}
\usepackage{mathtools}
\usepackage{xcolor}
\usepackage[english]{babel}

\makeatletter
\newcommand\RedeclareMathOperator{%
  \@ifstar{\def\rmo@s{m}\rmo@redeclare}{\def\rmo@s{o}\rmo@redeclare}%
}
\newcommand\rmo@redeclare[2]{%
  \begingroup \escapechar\m@ne\xdef\@gtempa{{\string#1}}\endgroup
  \expandafter\@ifundefined\@gtempa
     {\@latex@error{\noexpand#1undefined}\@ehc}%
     \relax
  \expandafter\rmo@declmathop\rmo@s{#1}{#2}}
\newcommand\rmo@declmathop[3]{%
  \DeclareRobustCommand{#2}{\qopname\newmcodes@#1{#3}}%
}
\@onlypreamble\RedeclareMathOperator
\makeatother

\DeclarePairedDelimiter\bra{\langle}{\rvert}
\DeclarePairedDelimiter\ket{\lvert}{\rangle}
\DeclarePairedDelimiterX\braket[2]{\langle}{\rangle}{#1 \delimsize\vert #2}

\newcommand{\ketPsi}[3][]{\ket{\Psi^{\mathrm{#1}}_{#2\mathbf{#3}}}}
\newcommand{\braPsi}[3][]{\bra{\Psi^{\mathrm{#1}}_{#2\mathbf{#3}}}}

\newcommand{\ketu}[3][]{\ket{u^{\mathrm{#1}}_{#2\mathbf{#3}}}}
\newcommand{\brau}[3][]{\bra{u^{\mathrm{#1}}_{#2\mathbf{#3}}}}
\newcommand{\braketu}[5][]{\braket{u^{\mathrm{#1}}_{#2\mathbf{#3}}}{{u^{\mathrm{#1}}_{#4\mathbf{#5}}}}}

\newcommand{\expi}[1]{\mathrm{e}^{\mathrm{i}#1}}
\newcommand{\exmi}[1]{\mathrm{e}^{-\mathrm{i}#1}}

\newcommand{\intBZdk}{\int\limits_{\mathrm{BZ}} \mathrm{d}\mathbf{k}}

\newcommand{\bk}{\mathbf{k}}
\newcommand{\bR}{\mathbf{R}}

\newcommand{\br}{\mathbf{r}}
\newcommand{\bb}{\mathbf{b}}
\newcommand{\bq}{\mathbf{q}}
\newcommand{\bD}{\mathbf{A}}
\newcommand{\bv}{\mathbf{v}}

\newcommand{\bZero}{\mathbf{0}}
\newcommand{\pR}{\mathcal{R}}

\DeclareMathOperator{\Tr}{Tr}
\RedeclareMathOperator{\Im}{Im}
\RedeclareMathOperator{\Re}{Re}

\begin{document}
	
\preprint{APS/123-QED}

\title{Self-consistent evaluation of the Berry connection for Wannier functions}

\author{Martin Th\"ummler}
\affiliation{%
	Institute of Physical Chemistry, Friedrich Schiller University Jena, Germany
}%
\author{Thomas Lettau}
\affiliation{
  Institute of Condensed Matter Theory and Optics, Friedrich Schiller University Jena, Germany 
}%
\author{Alexander Croy}%
\affiliation{%
	Institute of Physical Chemistry, Friedrich Schiller University Jena, Germany
}%
\author{Ulf Peschel}
\affiliation{
	Institute of Condensed Matter Theory and Optics and Abbe School of Photonics, Friedrich Schiller University Jena, Germany 
}%
\author{Stefanie Gr\"afe}
\affiliation{%
	Institute of Physical Chemistry, Abbe School of Photonics, Friedrich Schiller University Jena, Germany\\
	Fraunhofer Institute for Applied Optics and Precision Engineering, Jena, Germany
}%

\date{\today}

\begin{abstract}
The Berry connection is a gauge-dependent quantity frequently used to describe the optical response of solids.
Its evaluation requires a $\bk$-derivative with respect to the cell periodic-part of the Bloch-functions and is commonly calculated in the Wannier basis by using overlap matrices of cell-periodic parts of Bloch-functions at neighboring $\bk$-points. 
So far, all proposed interpolation schemes for the Berry connection do not account for the matrix structure of the overlap matrices explicitly but treat the matrix elements as independent, or only distinguish between diagonal and off-diagonal entries.
In this work, we propose a self-consistent interpolation scheme based on the matrix logarithm resulting in a strongly improved accuracy.
Furthermore, we discuss how the basis set incompleteness of the bands used in the ab-initio calculation imposes constraints on the accuracy.
We quantify the basis incompleteness based on the singular values of the overlap matrices and relate it to the invariant part of the spread functional $\Omega_{\mathrm{I}}$ of the Wannier functions.
Numerical calculations for monolayer $\mathrm{MoS}_2$ and bulk $\mathrm{Si}$ demonstrate that the proposed interpolation scheme is much less sensitive to the Wannierization details and leads to an improved quality of the velocity matrix and the optical conductivity.
\end{abstract}

\maketitle

\section{Introduction} \label{sec:intro}
Wannier functions \cite{wannier_structure_1937} are at the core of many ab-initio methods for solids, ranging from the description of polarization or magnetization \cite{noel_polarization_2001,thonhauserOrbitalMagnetizationPeriodic2005,lopez_wannier-based_2012,cohElectricPolarizationChern2009}, over topological effects \cite{wang_ab_2006,soluyanovWannierRepresentationZ22011,gresch_z2pack_2017}, to electron-phonon coupling \cite{giustino_electron-phonon_2007,lee_electronphonon_2023}.
Their localization in real space \cite{kohn_analytic_1959,monaco_optimal_2018} allows for an efficient interpolation of operator matrix elements from a coarse-grained ab-initio Monkhorst-Pack grid \cite{monkhors_1976} to a dense grid using discrete Fourier transforms \cite{marzari_1997,pizzi_2020,tsirkin_high_2021}.
This renders the numerical convergence with respect to the Brillouin zone (BZ) sampling feasible for many physical quantities \cite{marzari_2012,tsirkin_high_2021}.
The most commonly applied method for Wannierization results in maximally localized Wannier functions (MLWF) \cite{marzari_1997,marzari_2012} which is often combined with a band disentanglement \cite{souza_2001}, i.e., the smoothest possible reduction of the bands to the desired number of Wannier functions covering the energy range of interest.
Additional methods were developed to reliably find well localized Wannier functions \cite{damle_scdm-k_2017,damle2019variational} or to enforce symmetry constraints \cite{sakuma_symmetry-adapted_2013,koepernik_symmetry-conserving_2023,koretsune_construction_2023}.
The convergence of the Fourier-interpolated matrix elements with respect to the ab-initio grid size was extensively studied for $\bk$-local operators, e.g.\ for the Hamiltonian to obtain the band structure \cite{pizzi_2020}.

In this study, we consider and improve the accuracy of the non-local Berry connection, which recently attracted increasing interest \cite{lihm2026accurate,cole_exact_2026} due to its widespread use in computational material science.
Its matrix elements are also referred to as position or dipole matrix elements \cite{blount_formalisms_1962,marzari_1997} and are used in many applications \cite{xiao_berry_2010}, such as describing the polarization \cite{noel_polarization_2001,spaldin_beginners_2012} or optical responses \cite{ibanez-azpiroz_ab_2018,yates_spectral_2007,silva_high_2019,urru_optical_2025}. 
The Berry connection between different bands, $\bD^\bk_{mn} = \mathrm{i}\brau mk \nabla_\bk \ketu nk$, is numerically computed from the overlap matrices $M^{\bk\bb}_{mn}=\braket{ u_{m\bk} }{ u_{n{\bk + \bb}} }$ of the cell-periodic part of the Bloch functions with crystal momenta $\bk$ and $\bk+\bb$ of the coarse ab-initio grid \cite{marzari_1997}. 
In the initially proposed interpolation scheme \cite{wang_ab_2006}, the Berry connection is not Hermitian, which leads to nonphysical results, e.g.\ during the propagation of the semiconductor Bloch equations \cite{silva_high_2019}.
The translational invariance of the Wannier functions with respect to lattice vector shifts, manifests in the Berry connection as well \cite{blount_formalisms_1962,marzari_1997}.
It is respected for the diagonal matrix elements in the work of Marzari and Vanderbilt \cite{marzari_1997}, but not in the first work that Fourier-interpolates the Berry connection \cite{wang_ab_2006}.
Recently, Lihm developed a Hermitian scheme \cite{lihm2026accurate} that enforces the translational invariance also for the off-diagonal matrix elements to eliminate any spurious dependence of physical properties on the choice of the origin within the unit cell, e.g.\ for the optical conductivity \cite{lihm2026accurate}.
However, it is important not only to consider the origin shifts, but to ensure the correct transformation behavior of the Berry connection in general \cite{blount_formalisms_1962}.
An example is the consistency between different Wannierizations, like the MLWF and the separate Wannierization of the the conduction and valence bands (CB-VB) \cite{souza_2001,thummler_modeling_2026}.
Even recently proposed higher-order approximation schemes of the $\bk$-space derivative \cite{lihm2026accurate} provide no fundamental solution for that.

To address the aforementioned issues, we propose an interpolation scheme of the Berry connection $\bD^\bk$ that unifies the treatment of the diagonal and off-diagonal elements by accounting for the intrinsic matrix structure of the overlap matrices $M^{\bk\bb}$.
We identify $M^{\bk\bb}$ as the path-ordered matrix exponential of the Berry connection \cite{wilczek_appearance_1984}, which mixes all the matrix elements with each other.
This leads to an inverse problem, which we solve in two steps.
First, we take the matrix logarithm of the overlap matrices to compute an initial estimate for the Berry connection.
Second, we evaluate the path-ordered integrals based on the Fourier-interpolated estimate by employing a Magnus expansion and refine the Berry connection to reduce the error to the overlap matrices.
The second step is repeated until self-consistency is reached, which accelerates the convergence of the Berry connection with respect to the ab-initio grid size.
We note that the inherent incompleteness of the Wannier basis limits all interpolation schemes and will inevitable set a bound for the accuracy.
To asses its impact, we provide limits for the coupling of the Berry connection into the non-described part of the system based on the singular values of the overlap matrices $M^{\bk\bb}$ and discuss its close connection to the invariant part of the spread-functional $\Omega_{\mathrm{I}}$ introduced in Refs.\ \cite{marzari_1997,souza_2001}.

We assess the quality of the interpolation schemes based on the agreement between the ab-initio computed and the corresponding interpolated velocity operator.
The ab-initio velocity operator is an accurate reference as it is local in $\bk$ and can thus be interpolated as seamlessly as the Hamilton operator.
This allows for a momentum-resolved error estimation as well as a direct comparison of the different interpolation schemes without relying on convergence details of the Wannierization between different grid sizes.
This quality assessment is more systematic and transferable compared to approaches that consider the convergence of derived properties \cite{stengel_accurate_2006,cole_exact_2026}. 
In our numerical examples for monolayer $\mathrm{MoS_2}$ and bulk $\mathrm{Si}$, we demonstrate an improved accuracy of the velocity operator for our self-consistent logarithmic scheme.
Additionally, we show that optical conductivities computed based on the self-consistent logarithmic scheme are much more accurate and less sensitive to the Wannierization details compared to previous schemes, e.g.\ reductions of the relative deviations of the maximal optical conductivity from 28\% to less than 1\% for $\mathrm{MoS}_2$ on a $8\times8\times1$ ab-initio grid were achieved.

Our study is structured as follows:
Section~\ref{sec:theory} provides the theoretical background and introduces the the two predominantly used interpolation schemes for the Berry connection developed by Marzari and Vanderbilt, and Lihm, respectively.
We discuss these schemes in greater detail in Sec.~\ref{sec:consistentInterpolation} and introduce our (self-consistent) logarithmic interpolation scheme.
Afterwards, in Sec.~\ref{sec:results}, we quantify the the scheme-dependent error of the velocity operator, the basis incompleteness, and discuss their impact on computed optical conductivities for $\mathrm{MoS}_2$ and $\mathrm{Si}$.
Finally, we conclude in Sec.~\ref{sec:sao}.

\section{Theoretical background} \label{sec:theory}
In this section, we provide the theoretical background for the Wannier interpolation and review the existing interpolation schemes for the Berry connection.
We use atomic units throughout this manuscript if not otherwise indicated.

\subsection{Wannier interpolation}
We start with the Bloch functions $\ketPsi[H] nk$, the eigenfunctions of the cell-periodic Hamiltonian $\hat{H}$, where $n$ is the band index and $\bk$ the crystal momentum.
To improve readability, we do not consider spin explicitly, however, all presented derivations can be straight-forwardly generalized to include it.
The cell-periodic part of the Bloch-functions is $\ketu[H] nk = \exmi{\bk\hat{\br}} \ketPsi[H] nk$, where $\hat{\br}$ denotes the position operator.
The $\ketu[H] nk$ are generally not smooth in $\bk$, which hinders their interpolation to and the calculation of their derivative.
On the other hand, Wannier-functions $\ketPsi nk$ feature smooth cell-periodic parts $\ketu nk=\exmi{\bk\hat{\br}}\ketPsi nk$ due to their localization in the supercell \cite{marzari_2012,kohn_analytic_1959,monaco_optimal_2018}.
They are obtained from the Bloch-functions by applying semi-unitary transformation matrices $U^\bk$ as
\begin{align} \label{eq:BlochToWan}
    \ketPsi  nk = \sum\limits_{m} U^{\bk}_{mn} \ketPsi[H] mk
,
\end{align}
which are optimized to minimize a spread functional \cite{marzari_1997,souza_2001}.
The $n$th Wannier function (out of $n_{\mathrm{W}}$) in real space located at unit cell $\bR$ is given by a Fourier transform
\begin{align} \label{eq:defWF}
    \ket {\bR n} = \frac{1}{N_\bk}\sum\limits_{\bk} \exmi{\bk\bR} \ketPsi nk
,
\end{align}
where the $\bk$-sum is taken over the $N_\bk$ ab-initio grid points.
The Wannier functions allow for the interpolation of an arbitrary operator $\hat{O}$ in $\bk$-space, provided that it is local in $\bk$.
For that purpose, the Fourier transform of the matrix elements $O^\bk_{mn} = \bra{\Psi_{m\bk}} \hat{O} \ket{\Psi_{n\bk}}$ is calculated via
\begin{align} \label{eq:opKtoR}
    O^{\bR}_{mn} \equiv \bra {\bR m} \hat{O} \ket{\bZero n}= \frac{1}{N_\bk}\sum\limits_{\bk}  O^\bk_{mn} \expi{\bk\bR}
    .
\end{align}
The interpolation to an arbitrary $\bq$-point is done employing the inverse transform as
\begin{align} \label{eq:opRtoK}
    O^{\bq}_{mn} \equiv \bra{\Psi_{m\bq}} \hat{O} \ket{\Psi_{n\bq}}= \sum\limits_{\mathbf{R}} \bra {\bR m} \hat{O} \ket {\mathbf{0}n} \exmi{\bq\bR}
    ,
\end{align}
where the sum is taken over all the lattice vectors $\bR$ inside the Wigner-Seitz supercell \cite{pizzi_2020}.
For details we refer to Appendix~\ref{app:FTdetail}.
As we are interested in the interpolated Hamiltonian $\hat{H}$, we explicitly write the definition of its matrix elements in momentum space as
\begin{align}
    H^\bk_{mn} &= \braPsi mk \hat{H} \ketPsi nk\;.
\end{align}
The corresponding matrix elements $H^\bR_{mn}$ are defined via Eq.~\eqref{eq:opKtoR}.
Employing Eq.~\eqref{eq:opRtoK}, the analytical derivative of the Hamiltonian matrix elements reads
\begin{align}
    \nabla_{\bk} H^\bk_{mn} &= -\mathrm{i}\sum\limits_{\mathbf{R}} \bR H^{\bR}_{mn} \exmi{\bk\bR}\;.
\end{align}

\subsection{Definition and properties of the Berry connection}
The Berry connection \cite{blount_formalisms_1962,zak_berrys_1989} for a crystal momentum $\bk$ is most often defined for a single band.
In this manuscript, we denote the matrix elements between two states $m$ and $n$, defined as
\begin{align} \label{eq:defBerry}
    \bD^{\bk}_{mn} = \mathrm{i} \brau mk  \nabla_{\bk} \ketu nk
    ,
\end{align}
as the Berry connection. 
It is connected to the position operator via \cite{blount_formalisms_1962}
\begin{align} \label{eq:rk}
    \braPsi mk \hat{\br} \ketPsi nk= \bD^{\bk}_{mn} + \mathrm{i} \delta_{mn}\nabla_\bk
,
\end{align}
 where $\delta_{mn}$ is the Kronecker delta symbol.
The Berry connection is related to the velocity operator via the commutator relation $\hat{\bv} = \mathrm{i}[\hat{H}, \hat{\br}]$ as \cite{esteve-paredes_comprehensive_2023}
\begin{align} \label{eq:velocity}
    \bv^\bk = \mathrm{i} [ H^\bk, \bD^\bk] + \nabla_{\bk} H^{\bk}
.
\end{align}
We omitted the matrix indices in the above equation and will do so whenever convenient.
The Berry connection is Hermitian, i.e.\ $\bD^\bk = (\bD^\bk)^\dagger$, which follows directly from the orthogonality of the cell-periodic part of the Wannier functions, 
\begin{align} \label{eq:BerryHermiticity}
    0 = \nabla_{\bk} \braketu mknk = \mathrm{i} \left[(\bD^{\bk}_{nm})^{*} - \bD^{\bk}_{mn}  \right]
.
\end{align}
Its Fourier transform
\begin{align}
    \bD^\bR =  \frac{1}{N_\bk} \sum\limits_{\bk} \bD^\bk \expi{\bk\bR}\;,
\end{align}
is evaluated as
\begin{align} \label{eq:BerryR}
    \bD^\bR_{mn} &= \bra{\bR m} \hat{\br} \ket{\bZero n}
.
\end{align}
The $\bD^\bR_{mn}$ are named position operator matrix elements as well \cite{marzari_1997}.
The Berry connection is gauge dependent, i.e., a $\bk$-dependent unitary basis transformation $U^\bk$ of the cell-periodic part of the Wannier functions 
\begin{align} \label{eq:defTrafo}
    \ketu nk \rightarrow \sum\limits_{m} U^\bk_{mn} \ketu mk
\end{align}
leads to \cite{blount_formalisms_1962}
\begin{align} \label{eq:BerryTrafo}
	\bD^{\bk} \rightarrow U^{\bk\dagger} \bD^\bk U^\bk + \mathrm{i} U^{\bk\dagger} \nabla_\bk U^\bk
.
\end{align}
Important special cases of these gauge transformations are permutations of the band indices or shifts of the Wannier functions via 
\begin{align} \label{eq:BerryTI}
    U^{\bk}_{nm} = \delta_{mn} \exmi{\bk\pR_n}
\end{align}
with shift vectors $\pR_{n}$.
If all the $\pR_{n}$ are lattice vectors, Eq.~\eqref{eq:BerryTrafo} reduces to the translational invariance condition:
The physics must be independent of the choice of the origin and the unit cell to which the Wannier functions are localized.
This condition is for instance accounted for in the in the definition of the Wannier function spread, see Eqs.\ (31), (32), and (36) in Ref.\cite{marzari_1997}.

\subsection{Review of existing interpolation schemes for the Berry connection}
In this section, we review the commonly used schemes to obtain and interpolate the Berry connection from ab-initio calculated Wannier functions.
In all the interpolation schemes, the following two steps are performed:
First, the $\bD^\bk$ (or equivalently the $\bD^\bR$) matrices are calculated on the ab-initio $\bk$-grid (of the supercell) and 
second, the interpolation to an arbitrary $\bq$-point is done by employing Eq.~\eqref{eq:opRtoK}, which relies on the smoothness of the $\ketu nk$ in momentum space \cite{wang_ab_2006}.
The interpolation schemes differ only in the calculation of the $\bD^\bk$ ($\bD^\bR$) matrices from the overlap matrices
\begin{align} \label{eq:defMkb}
    M^{\bk\bb}_{mn} = \braket{ u_{m\bk} }{ u_{n{\bk + \bb}} }
    ,
\end{align}
where $\bk$ and $\bk + \bb$ are on the ab-initio grid.
For the real space discussion, we define the Fourier-transform of the overlap matrices as
\begin{align} \label{eq:defMR}
    M^{\bR\bb} = \frac{1}{N_\bk}\sum\limits_{\bk} M^{\bk\bb}  \expi{\bk\bR}
\end{align}
and obtain \cite{marzari_1997}
\begin{align} \label{eq:MR}
    M^{\bR\bb}_{mn} = \bra{\bR m} \exmi{\bb\hat{\br}} \ket{\bZero n}
,
\end{align}
which follows from the definition of the Wannier functions, see Eq.~\eqref{eq:defWF}.
In the following, we consider all the equations for supercell sizes approaching infinity, which enables us treat the $\bk$'s as continuous variables.
The Berry connection, see Eq.~\eqref{eq:defBerry}, is then given in the limit $\bb \rightarrow \bZero$ by
\begin{align}
    \bD^{\bk} = \mathrm{i} \nabla_{\bb} M^{\bk\bb}
.
\end{align}
Combining Eqs.~\eqref{eq:MR} and \eqref{eq:BerryR} leads for the position operator matrix elements to
\begin{align} \label{eq:BerryMRb}
    \bD^\bR = \mathrm{i} \nabla_{\bb} M^{\bR\bb}
.
\end{align}
On the ab-initio grid, we compute the gradient of a $\bk$-dependent function $f$ by employing finite differences
\begin{align} \label{eq:gradientError}
	\nabla_\bk f(\bk) = \sum_{\bb} w_{\bb} \bb f(\bk + \bb) + \mathcal{O}(|\bb|^{2})
\end{align}
with the Euclidean norm $|\cdot|$. 
The weights $w_{\bb}$ are chosen such that they obey the condition
\begin{align} \label{eq:bSum}
	\sum\limits_{\bb} w_{\bb} b_\alpha b_\beta = \delta_{\alpha\beta}
	,
\end{align}
and are symmetric
\begin{align} \label{eq:weightCondition}
	w_{\bb} = w_{-\bb} \text{ and } w_{\bZero} = 0\;.
\end{align}
Here, $b_\alpha$ and $b_\beta$ are the real space components of $\bb$ in $\alpha$ and $\beta$ direction, respectively \cite{marzari_1997}.
The symmetry condition on the weights in Eq.~\eqref{eq:weightCondition} leads to quadratic error scaling in Eq.~\eqref{eq:gradientError}.
Even if the condition in Eq.~\eqref{eq:weightCondition} is not strictly necessary, it is still satisfied in Ref.\ \cite{marzari_1997,marzari_2012} and will later help to enforce the Hermiticity of the interpolation schemes.
Applying Eq.~\eqref{eq:gradientError} to the off-diagonal elements ($m\neq n$) of $M^{\bk\bb}$  \cite{wang_ab_2006}
leads to
\begin{align} \label{eq:MV}
	\bD^{\mathrm{MV},\bk}_{mn} = \mathrm{i} \sum\limits_{\bb} w_\bb \bb M^{\bk\bb}_{mn}
,
\end{align}
where we introduced the superscript MV denoting Marzari and Vanderbilt.
However, the diagonal elements must respect the translational invariance condition, see Eqs.~\eqref{eq:BerryTrafo} and \eqref{eq:BerryTI}.
This is ensured by employing the expansion $\ln(1+x) = x + \mathcal{O}(x^2)$ for the diagonal elements, leading to
\begin{align} \label{eq:MVdiag}
	\bD^{\mathrm{MV},\bk}_{nn} = - \Im\sum\limits_{\bb} w_{\bb} \bb \ln M^{\bk\bb}_{nn}
.
\end{align}
The Fourier transform of the Berry connection, see Eq.~\eqref{eq:BerryR}, allows to define the Wannier centers as
\begin{align}
	\br_n  \equiv \bD^{\bR=\bZero}_{nn}
\end{align}
yielding
\begin{align} \label{eq:MVcenter}
	\br^{\mathrm{MV}}_{n} = -\frac{1}{N_\bk}\Im\sum\limits_{\bk\bb} w_{\bb} \bb \ln M^{\bk\bb}_{nn}
.
\end{align}
Lihm proposed a Hermitian scheme for the off-diagonal elements of the Berry connection \cite{lihm2026accurate}, by enforcing the translational invariance of the off-diagonal position operator matrix elements.
Its implementation is provided in \cite{tsirkin_high_2021}.
Within this approach, one evaluates Eq.~\eqref{eq:BerryMRb} by a Taylor expansion of the exponential phase factor in $\bra{\bR m} \exmi{\bb\hat{\br}} \ket{\bZero n}$ with respect to $\bb$ around the midpoint of the two Wannier function centers.
This reduces the expansion error of the position operator in the real-space regions of the supercell where the Wannier functions overlap most.
The scheme reads for $m\neq n$ or $\bR \neq \bZero$ 
\begin{align} \label{eq:LihmR}
    \bD^{\mathrm{Lihm},\bR}_{mn} = \mathrm{i} \sum\limits_{\bb} w_\bb \bb M^{\bR\bb}_{mn}\expi{\frac{\bb}{2}(\br_n + \br_m-\bR)}
.
\end{align}
The diagonal elements $\br_n^\mathrm{Lihm}=\bD^{\bR=0}_{nn}$ are obtained from an expression of the Wannier centers that respects their transformation behavior under origin shifts such as Eq.~\eqref{eq:MVcenter} or similar formulations \cite{resta_quantum-mechanical_1998,stengel_accurate_2006}.
In addition, Lihm refines the Wannier centers recursively until
\begin{align} \label{eq:LihmCenter}
    \br_n = \frac{\mathrm{i}}{N_\bk}\sum\limits_{\bk\bb} w_{\bb} \bb M^{\bk\bb}_{nn} \expi{\bb\br_n}
\end{align}
holds.
Transforming Eq.~\eqref{eq:LihmR} into momentum space by accounting for the Wannier centers yields
\begin{align} \label{eq:Lihm}
    \bD^{\mathrm{Lihm},\bk}_{mn} =& \mathrm{i} \sum\limits_{\bb} w_\bb \bb M^{\bk-\frac{\bb}{2},\bb}_{mn}\expi{\frac{\bb}{2}(\br_n + \br_m)}
       \nonumber \\
       &\,+\delta_{mn} \left( \br_n - \frac{\mathrm{i}}{N_\bk}\sum\limits_{\bk'\bb} w_{\bb} \bb M^{\bk'\bb}_{nn} \expi{\bb\br_n}\right)
,
\end{align}
where $M^{\bk-\frac{\bb}{2},\bb}_{mn}$ indicate the Fourier-interpolated overlap matrix elements.

\section{Self-consistent interpolation scheme} \label{sec:consistentInterpolation}
In this section, we evaluate the existing interpolation schemes for the Berry connection and develop our self-consistent logarithmic scheme step by step.
We show first that the MV-scheme is not Hermitian and propose a symmetric (sym) scheme as remedy.
Next, we discuss the close connection of the Lihm-scheme to the sym-scheme.
Afterwards, we present the logarithmic (log) scheme by accounting for the matrix structure of the overlap matrix properly and introduce the self-consistent logarithmic (sclog) scheme by extending the log-scheme from a local to a global interpolation scheme.
Lastly, we discuss the errors introduced by assuming the basis set completeness required to justify the (sc)log-scheme.

\subsection{Hermiticity enforcement} \label{sec:MVsym}
Testing the MV interpolation scheme, see Eq.~\eqref{eq:MV}, for its Hermiticity yields ($n\neq m$)
\begin{align} \label{eq:HermiticityCheck}
    (\bD^{\mathrm{MV},\bk, \dagger})_{mn} &= -\mathrm{i} \sum\limits_{\bb} w_\bb \bb M^{\bk+\bb,-\bb}_{mn} \nonumber \\
    &= \mathrm{i} \sum\limits_{\bb} w_\bb \bb M^{\bk-\bb,\bb}_{mn}\;,
\end{align}
where we used $(M^{\bk\bb})^\dagger = M^{\bk+\bb,-\bb}$ and Eq.\ \eqref{eq:weightCondition}.
As the $M^{\bk\bb}$ matrices change throughout the BZ, the Hermiticity is generally not given, compare to Eq.~\eqref{eq:MV}.
However, by using the the Fourier-interpolated $M^{\bk-\frac{\bb}{2},\bb}$ matrices instead, which obey
\begin{align} \label{eq:MdaggerSym}
    (M^{\bk-\frac{\bb}{2},\bb})^\dagger = M^{\bk+\frac{\bb}{2},-\bb}
    ,
\end{align}
see Appendix~\ref{app:Mdagger}, the Hermiticity is restored. 
Thus, we propose to calculate the off-diagonal elements of the Berry connection ($n\neq m$) as 
\begin{align} \label{eq:sym}
    \bD^{\mathrm{sym},\bk}_{mn} &= \mathrm{i} \sum\limits_{\bb} w_\bb \bb M^{\bk-\frac{\bb}{2},\bb}_{mn}
.
\end{align}
The Hermiticity of $\bD^{\mathrm{sym},\bk}$ is confirmed by the same calculation as in  Eq.~\eqref{eq:HermiticityCheck}.
We now show, that this symmetric interpolation scheme is consistent, i.e., it converges to the Berry connection for $\bb \rightarrow \bZero$, by evaluating the Taylor expansion of $M^{\bq\bb}$ around $\bq=\bk+\frac{\bb}{2}$ as
\begin{align} \label{eq:MkTaylorSym}
    M^{\bq\bb}_{mn} &= \braket{u_{m\bq-\frac{\bb}{2}}}{u_{n\bq+\frac{\bb}{2}}}
    \nonumber \\
    &=
    \delta_{mn} + \frac{\bb}{2}\left[
        \braket{ u_{m\bq}}{ \nabla_\bk u_{n\bq}}-
         \braket{\nabla_{\bq}u_{m\bq}}{u_{n\bq}} 
     \right]   
     \nonumber \\ &\quad
     +
     \mathcal{O}(|\bb|^2) \nonumber \\
    &= \delta_{mn} - \mathrm{i} \bb\bD^{\bk+\frac{\bb}{2}}_{mn} + \mathcal{O}(|\bb|^2)
,
\end{align}
where we used Eq.~\eqref{eq:BerryHermiticity} in the last line.
Interestingly, $\bD^{\mathrm{sym},\bk}$ is accurate to second order in $\bb$, see the discussion after Eq.~\eqref{eq:weightCondition}.
For the diagonal elements, we employ the MV formula, i.e.,
\begin{align}
    \bD^{\mathrm{sym},\bk}_{nn} \equiv \bD^{\mathrm{MV},\bk}_{nn}
.
\end{align}

\subsection{Lihm's scheme} \label{sec:Lihm}
We first check the diagonal elements of $\bD^{\mathrm{Lihm,\bk}}$ for their consistency.
Rewriting the weighted sums over $\bb$ in Eq.~\eqref{eq:Lihm} as gradient yields
\begin{align}
    \bD^{\mathrm{Lihm},\mathbf{k}}_{nn} &= \mathrm{i}\nabla_\bb\left[M_{nn}^{\bk-\frac{\bb}{2},\bb}\expi{\bb \br_n}\right]  \nonumber \\
                                          &\quad + \br_n - \frac{\mathrm{i}}{N_\bk}\nabla_\bb\sum\limits_{\bk'}\left[M_{nn}^{\bk'\bb}\expi{\bb\br_n}\right]
,
\end{align}
which simplifies to
\begin{align} \label{eq:LihmDiag}
    \bD^{\mathrm{Lihm},\bk}_{nn} = \mathrm{i} \nabla_{\bb}M_{nn}^{\bk-\frac{\bb}{2},\bb} + \br_n - \frac{\mathrm{i}}{N_\bk}\nabla_\bb \sum\limits_{\bk'} M^{\bk'\bb}_{nn}
.
\end{align}
The first term describes the Berry connection, compare to Eq.~\eqref{eq:MkTaylorSym}.
Therefore the remaining terms must add up to zero for $\bb \rightarrow \bZero$ to obtain a locally consistent interpolation scheme, i.e.,
\begin{align}
    \br_{n}= \frac{\mathrm{i}}{N_\bk}\nabla_\bb \sum\limits_{\bk} M^{\bk\bb}_{nn} + \mathcal{O}(|\bb|)
.
\end{align}
This condition is satisfied by the MV Wannier center, see Eq.~\eqref{eq:MVcenter}, with a quadratic error scaling because of Eq.~\eqref{eq:weightCondition} and holds for the Lihm Wannier centers in Eq.~\eqref{eq:LihmCenter} as well.
For the off-diagonal matrix elements, Lihm's scheme, see Eq.~\eqref{eq:Lihm}, is very similar to the symmetric scheme, see Eq.~\eqref{eq:sym}, but contains additional phase factors of $\expi{\frac{\bb}{2}(\br_m+\br_n)}$.
It is straight-forward to show that these phase factors arise when evaluating Eq.~\eqref{eq:sym} for all $\br_n=\bZero$ and shifting the Wannier centers by using $U^\bk_{mn} = \delta_{mn}\exmi{\bk\br_n}$ in Eq.~\eqref{eq:defTrafo}. 
Equation~\eqref{eq:LihmCenter}, which is used to compute the Lihm Wannier centers, ensures the equivalence to the symmetric scheme for diagonal elements.
Thus, Lihm's scheme is identical to the symmetric scheme evaluated in a specific basis.

\subsection{Matrix logarithm} \label{sec:log}
Up to now, all the presented interpolation schemes treat the matrix elements independently (for different Wannier indices) of each other, or distinguish only between diagonal and off-diagonal matrix elements.
However, as the matrices $M^{\bk\bb}$ connect two orthonormal basis sets, we expect them to generate an unitary transformation and hence propose to use the matrix logarithm to compute $\bD^\bk$ instead of taking the logarithm of the diagonal matrix elements.
More precisely, we  write the $M^{\bk\bb}$ as a result of a path-ordered matrix exponential \cite{wilczek_appearance_1984}
\begin{align} \label{eq:BerryGenerator}
    M^{\bk\bb} = \mathcal{T}\exp\left(-\mathrm{i}\int\limits_{0}^{|\bb|} \bD^{\bk+t\frac{\bb}{|\bb|}} \mathrm{d}t \right)
\end{align}
in a symbolic notation, where $\mathcal{T}$ is the path-ordering operator.
In this sense, $\bD^\bk$ is the generator for the $M^{\bk\bb}$ matrices, in contrast to the previous interpolation schemes, which assume incorrectly a purely element-wise relation between the matrix elements of $M^{\bk\bb}$ and $\bD^\bk$.
We note that this representation is only exact for a complete basis as we inserted resolutions of identity during its derivation.
However, the basis must be nearly complete in the energy region of interest regardless of that to obtain converged calculations.
We comment on the basis incompleteness further in Sec.~\ref{sec:basisSpill}.
Taking the logarithm of Eq.~\eqref{eq:BerryGenerator} and employing the Magnus expansion yields \cite{magnus_exponential_1954,blanes_magnus_2009}
\begin{align} \label{eq:logM}
    \log M^{\bk\bb} = -\mathrm{i} \bb \bD^{\bk+\frac{\bb}{2}} + \mathcal{O}(|\bb|^3)\;.
\end{align}
The matrix elements of the Berry connection and the overlap matrices are evaluated at different $\bk$-points in Eq.~\eqref{eq:logM}, which requires an interpolation of either the $M^{\bk\bb}$ matrices or of their logarithm.
In order to avoid an interpolation of the $M^{\bk\bb}$ matrices, we take the matrix logarithm first and interpolate afterwards, i.e.,
\begin{align} \label{eq:log}
    \bD^{\mathrm{log},\bR} = \frac{\mathrm{i}}{N_\bk} \sum\limits_{\bk\bb} w_\bb \bb \expi{\left(\bk-\frac{\bb}{2}\right)\bR}\log M^{\bk\bb}
,
\end{align}
where it is sufficient to rely on the principal branch of the matrix logarithm when $M^{\bk\bb}$ is close to unity.
In some cases, the matrix logarithm may be close to a branch cut, e.g. when the BZ is sampled very coarsely or only at the $\Gamma$-point.
In Appendix~\ref{app:branch}, we show how to use the guiding centers defined for the Wannierization procedure to avoid any inconsistent branch selection.
In case of a complete basis, the Hermiticity of this scheme follows directly from the anti-Hermiticity of $\log M^{\bk\bb}$ without relying on $w_\bb = w_{-\bb}$.
This argument is not applicable to the other interpolation schemes.
The logarithmic scheme is, however, still dependent on the choice of the origin.
In Appendix~\ref{app:altLogScheme}, we discuss the alternative interpolation scheme, where we first interpolate the overlap matrices before taking the logarithm to ensure its independence of the origin choice.
In our numerical studies, this alternative scheme performed slightly worse than the logarithmic scheme.
This behavior was also found for its self-consistent counterpart presented in the next section.
In Appendix~\ref{app:sizeConsistent}, we show that for both schemes the computed polarization is size-consistent in the sense of Ref.~\cite{stengel_accurate_2006}.

\subsection{Self-consistent interpolation} \label{sec:clog}
All interpolation schemes presented so far are local in $\bk$ and have a quadratic error scaling in $|\bb|$.
It has been proposed to enhance the accuracy of the gradient approximation for the Berry connection by using a higher-order interpolation scheme for Eq.~\eqref{eq:gradientError} \cite{lihm2026accurate}. 
However, this is not required for our self-consistent logarithmic (sclog) interpolation scheme:
After computing the Berry connection locally for every grid point via the logarithmic scheme, we evaluate the path-ordered integrals, see Eq.~\eqref{eq:BerryGenerator}, explicitly by employing the global Fourier interpolation of the estimated Berry connection.
This gives us the deviation to the $M^{\bk\bb}$ matrices, which we minimize by recursively adapting the $\bD^\bk$ matrices.
We define
\begin{align}
    S^{\bk}_{\bb} = -\mathrm{i}\bb\bD^{\bk+\frac{\bb}{2}}
\end{align}
and evaluate the path ordered product by a Magnus-expansion of fourth order \cite[Eq.~(256)]{blanes_magnus_2009} as
\begin{align} \label{eq:MagnusExpansionInterpol}
    I^{\bk\bb} &= \log \mathcal{T}\exp\left(-\mathrm{i}\int\limits_{0}^{|\bb|} \bD^{\bk+t\frac{\bb}{|\bb|}} \mathrm{d}t \right)
    \nonumber \\
               &\approx \frac{S^{\bk-\frac{\bb}{2}}_\bb + 4 S^{\bk}_\bb + S^{\bk+\frac{\bb}{2}}_\bb}{6} - \frac{ [S^{\bk-\frac{\bb}{2}}_\bb, S^{\bk+\frac{\bb}{2}}_\bb]}{12}
               ,
\end{align}
where the $S^{\bk-\frac{\bb}{2}}_\bb$ and $S^{\bk+\frac{\bb}{2}}_\bb$ are obtained via Fourier interpolation.
We apply the recursion
\begin{align}
    S^\bk_{\bb,0} &= \log M^{\bk\bb}\;,
    \nonumber \\
    \label{eq:clogRecursion}
    S^{\bk}_{\bb,n+1} &= S^\bk_{\bb,n} + \log M^{\bk\bb} - I^{\bk\bb}_n
,
\end{align}
until we reach the limit
\begin{align}
    S^{\mathrm{sclog}, \bk}_\bb = \lim\limits_{n\to \infty} S^{\bk}_{\bb, n}
.
\end{align}
Afterwards, we compute the self-consistent Berry connection as
\begin{align} \label{eq:clog}
    \bD^{\mathrm{sclog},\bR} = \frac{\mathrm{i}}{N_k} \sum\limits_\bb w_\bb \bb \expi{\left(\bk-\frac{\bb}{2}\right)\bR} S^{\mathrm{sclog},\bk}_\bb
,
\end{align}
which minimizes the error globally.
The local error scales at least as $\mathcal{O}(|\bb|^4)$, as given by the order of the Magnus expansion.
In our numerical experiments higher-order Magnus expansions did not improve the results.
We remark that the convergence criterion for the Magnus series,
\begin{align}
    \int\limits_{0}^{|\bb|} \Vert\bD^{\bk+t\frac{\bb}{|\bb|}}\Vert \mathrm{d}t \approx |\bb|\,\Vert\bD^{\bk+\frac{\bb}{2}}\Vert < \pi
\end{align}
with the Frobenius norm $\Vert\cdot\Vert$ \cite{moan_convergence_2008}, is satisfied even for coarse grid sizes as $|\bb| \Vert\bD^{\bk+\frac{\bb}{2}}\Vert = \Vert M^{\bk\bb} - \mathbb{1}\Vert + \mathcal{O}(|\bb|^3)$.
An improved consistency through the recursive refinement is expected for sufficiently large supercells that can capture the behavior of the Wannier functions, which is a fundamental constraint to the Wannier interpolation itself.
We emphasize that the $S^{\mathrm{sclog}, \bk}_\bb$ matrices are not Hermitian as long as the $M^{\bk\bb}$ matrices are not unitary.
However, the sclog-scheme is Hermitian since during all iteration steps $S^{\bk\dagger}_{\bb,n} = S^{\bk+\bb}_{-\bb,n}$ holds and because of $w_\bb = w_{-\bb}$ in Eq.~\eqref{eq:clog}.
We also tested an analogous scheme that enforces the self-consistency for the Berry connection $\bD^\bk$ (instead of the $S^\bk_\bb$) matrices at each iteration step.
However, this led to worse results, which we attribute to the additional misalignment induced by the enforced unitarity of the Magnus expansion in Eq.~\eqref{eq:MagnusExpansionInterpol} at each iteration step that is not compatible with the non-unitarity of the $M^{\bk\bb}$ matrices, see Appendix~\ref{app:scConvergence} for numerical details.

There are two computational aspects we comment on:
First, no numerical issues to compute the matrix logarithm are expected as it is an implicit prerequisite of the Wannierization procedure that the overlap matrices are never close to being singular and are explicitly optimized for in the disentanglement procedure.
Second, the overall computational effort is dominated by computing the matrix logarithm and therefore larger than for the previous schemes, but still much lower than the effort to compute the Wannierization itself. 
The overhead due to the fix-point iteration is marginal as fast Fourier transforms can be employed to compute the $I^{\bk\bb}$ matrices.

\subsection{Quantifying the basis set incompleteness} \label{sec:basisSpill}
In numerical calculations, the number of Wannier functions is finite and they typically do not form a complete basis set.
We express the augmentation of the overlap matrices to a compete basis as
\begin{align} \label{eq:MkComplete}
    M^{\mathrm{complete},\bk\bb} &= 
    \begin{pmatrix}
        M^{\bk\bb} & \tilde{M}^{\mathrm{spill},\bk\bb} \\
        M^{\mathrm{spill},\bk\bb} & M^{\mathrm{off},\bk\bb} 
    \end{pmatrix} \\
    &=
    \mathbb{1} - \mathrm{i} \bb
    \begin{pmatrix}
        \bD^{\bk} & \left(\bD^{\mathrm{spill},\bk}\right)^\dagger \\
        \bD^{\mathrm{spill},\bk} & \bD^{\mathrm{off},\bk}  
    \end{pmatrix}
    + \mathcal{O}(|\bb|^2) \nonumber 
    ,
\end{align}
where the superscript 'spill' describes the coupling to the non-described part denoted by 'off'.
To calculate bounds for the spill $\bD^{\mathrm{spill},\bk}\mathbf{e}_\bb$ in direction $\mathbf{e}_{\bb} = \bb/|\bb|$, we denote the singular values of $M^{\bk\bb}$ as $\sigma^{\bk\bb}_1, \dots, \sigma^{\bk\bb}_{n_\mathrm{W}}$ in decreasing order.
Because of the sine-cosine decomposition \cite{golub2013matrix}, the singular values of $M^{\mathrm{spill},\bk\bb}$ are given by
\begin{align} \label{eq:sigmaSpill}
    \sigma^{\mathrm{spill},\bk\bb}_n = \sqrt{1 - (\sigma^{\bk\bb}_{1+n_\mathrm{W}-n})^2 }
\end{align}
in decreasing order.
For $\bb\rightarrow \bZero$, the infinitesimal spill is bounded by 
\begin{align} \label{eq:infinitesimalSpill}
    \frac{\sigma^{\mathrm{spill},\bk\bb}_{n_{\mathrm{W}}}}{|\bb|} \lesssim |\bD^{\mathrm{spill},\bk}\mathbf{e}_{\bb}| \lesssim \frac{\sigma^{\mathrm{spill},\bk\bb}_{1}}{|\bb|}
\end{align}
by means of perturbation theory as we show in Appendix~\ref{app:basisSpill}.
Furthermore, we relate it in Appendix~\ref{app:basisSpill} to the invariant part of the Wannier function spread $\Omega_{\mathrm{I}}$ \cite{marzari_1997,souza_2001} as
\begin{align} \label{eq:localSpillBZ}
 \Omega_{\mathrm{I}}^V &\equiv \frac{(2\pi)^3}{V} \Omega_{\mathrm{I}}
 \nonumber \\
  &= \sum\limits_{\mathbf{e}_{\bb}} w_{\mathbf{e}_{\bb}}  \intBZdk \Tr\left\{\mathbf{e}_\bb\bD^{\mathrm{spill},\bk}(\mathbf{e}_\bb\bD^{\mathrm{spill},\bk})^\dagger\right\}
\end{align}
with an error scaling of $\mathcal{O}(|\bb|^2)$, where $V$ is the unit cell volume and $w_{\mathbf{e}_{\bb}}= w_\bb|\bb|^2$.
The quantity $\Omega_{\mathrm{I}}^V$ represents a directionally weighted measure of the basis spill integrated over the BZ and is thus comparable between different systems.
The disentanglement procedure of Souza \cite{souza_2001} minimizes $\Omega_{\mathrm{I}}$ to obtain a reduced number of smooth bands within a certain energy range.
As a consequence, it selects the Wannier functions such that the basis spill effects occurring in the log- and sclog-scheme are minimized.
We note that the non-zero value of $\Omega_{\mathrm{I}}^V$ is an inherent property of the finite number of the Wannier functions.

In the log- and sclog-scheme, the matrix logarithm is computed from the $M^{\bk\bb}$ and not from the $M^{\mathrm{complete},\bk\bb}$ matrices.
The logarithm is therefore only accurate to $\mathcal{O}(|\bb|^2)$, which can be seen by employing a Taylor-expansion.
The leading error coefficient is determined by the basis incompleteness and does not change when the Wannierization is adapted as long as the disentanglement, i.e., the spanned subspace, is not modified.

\section{Numerical Results} \label{sec:results}
We test the interpolation schemes numerically for monolayer $\mathrm{MoS}_2$ and bulk $\mathrm{Si}$ for two different Wannierizations and different ab-initio grid sizes and demonstrate that the schemes behave very differently, with the log- and sclog-scheme performing best.
First, we discuss the mismatch between the scheme-dependent and the ab-initio computed velocity operator qualitatively as well as quantitatively.
Afterwards, we quantify the basis incompleteness and discuss its convergence with respect to the ab-initio grid size.
Lastly, we demonstrate the serious impact of the interpolation schemes on optical conductivities.

\subsection{Parametrization and implementation details}
All density functional theory (DFT) calculations are performed with Quantum Espresso (QE) 7.5 \cite{giannozzi_2009,giannozzi_2017,giannozzi_2020}.
We obtain the scheme-dependent velocity matrix elements via Eq.~\eqref{eq:velocity}.
For the ab-initio matrix elements $\bv^{\mathrm{abi},\bk}_{mn} = \braPsi mk \hat\bv \ketPsi nk$, we patched pw2wannier90.x inside of QE based on the formulas of Refs.~\cite{kageshima_momentum-matrix-element_1997,tobik_electric_2004} including the correction due to the non-local pseudopotentials.
The Lihm scheme is evaluated with the centers from Eq.~\eqref{eq:LihmCenter}.
We compute the matrix logarithm required for the log- and sclog-scheme using the algorithm in Ref.~\cite{al-mohy_improved_2012} extended by an branch selection routine as outlined in Appendix~\ref{app:branch}.
During the evaluation of the sclog-scheme, we performed 20 iterations of Eq.~\eqref{eq:clogRecursion} to safely reach its numerical convergence, see Appendix~\ref{app:scConvergence} for details.
Our code and the input files for the DFT calculations, Wannierizations and the evaluation of the Berry connection schemes are available at \cite{thuemmler_2026}.

For monolayer $\mathrm{MoS}_2$, we employed the Perdew-Burke-Ernzerhof functional \cite{perdew_1996} and the pseudopotential from \cite{hamann_2013} with a plane-wave energy-cutoff of 90\,Ry in two dimensions \cite{sohier_density_2017} to obtain the self-consistent ground state density on a $9\times9\times1$ $\bk$-grid.
Using Wannier90 \cite{marzari_2012}, we then computed two Wannierzations for each of the $N_k\times N_k\times 1$ grids with 11 bands.
We use the maximally localized Wannier functions (MLWFs) of all bands as the first Wannierization.
The second one is obtained by stitching the separate Wannierizations of the valence and conduction bands (CB-VB) together.
Band disentanglement \cite{souza_2001} was not required since the selected bands are energetically separated from the remaining ones.
In the following, we restrict all evaluations for monolayer $\mathrm{MoS}_2$ to the in-plane components.
The out-of-plane components and the importance of the guided branch cut selection of the matrix logarithm is discussed in Appendix~\ref{app:branch}.

\begin{figure}[tb]
    \centering
    \includegraphics[width=\columnwidth]{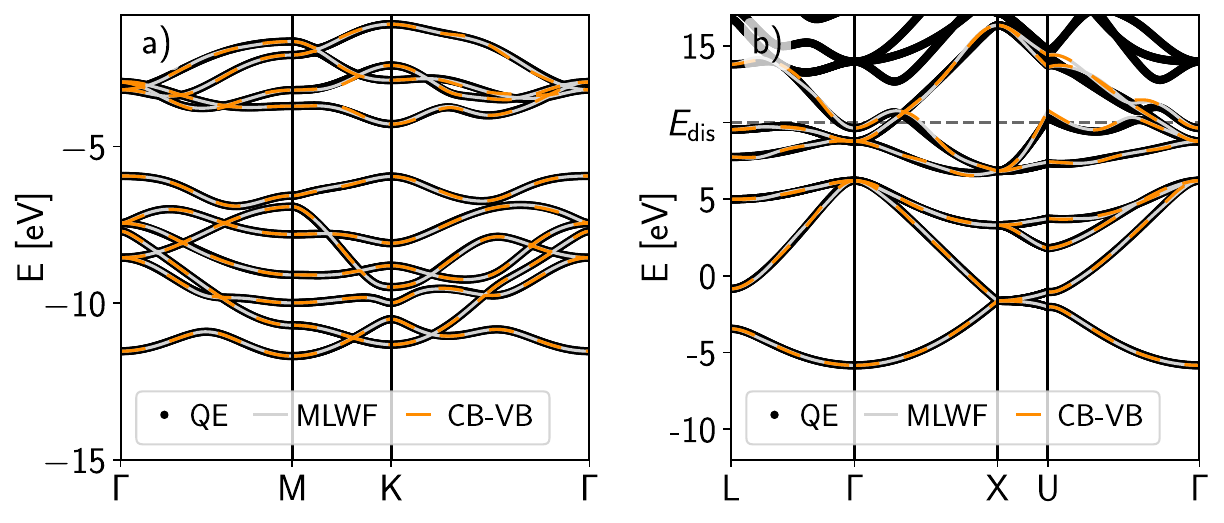}
    \caption{Band structure for a) monolayer $\mathrm{MoS}_2$ ($6 \times 6\times1$ $\bk$-grid) and b) bulk silicon ($6 \times 6 \times 6$ $\bk$-grid).
    The DFT calculation performed with Quantum Espresso is depicted with thick black dots.
    The gray and orange lines denote the Wannier-interpolated band structure based on the MLWF and CB-VB Wannierizations, respectively.
    The dashed line in b) depicts the upper limit of the frozen window used for the disentanglement of the Wannier functions.
    }
    \label{fig:bandstructure}
\end{figure}
Similarly, we used the pseudopotential of \cite{von_barth_1980} for $\mathrm{Si}$ with the local density approximation and a cutoff of 100\,Ry to compute the ground state density on a $12\times12\times12$ grid.
We obtained the MLWF and CB-VB Wannierization of 8 bands for each grid size $N_k\times N_k \times N_k$ with the upper end of the frozen disentanglement window at $E_{\mathrm{dis}}=10\,\mathrm{eV}$ \cite{souza_2001}.
The corresponding band structures obtained from the Wannierizations (with $N_k=6$) for $\mathrm{MoS}_2$ and $\mathrm{Si}$ are shown in Fig.~\ref{fig:bandstructure}, which agree very well with band structure that was calculated with QE.

\subsection{Scheme-dependent accuracy of the velocity operator} \label{sec:comparisonVM}
\begin{figure}[tb]
    \centering
    \includegraphics[width=\columnwidth]{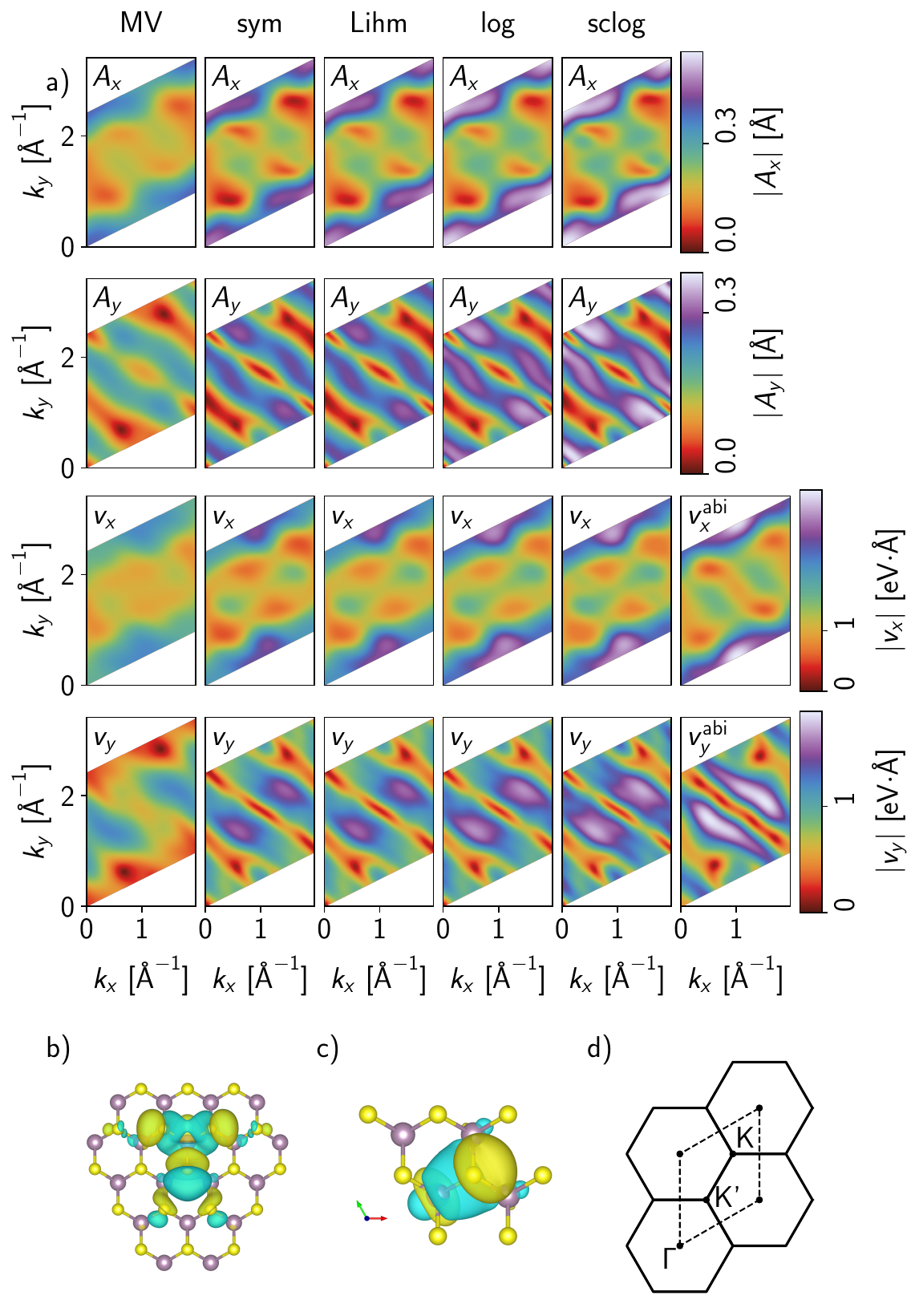}
    \caption{a) BZ-resolved magnitude of the Berry connections $A_{x/y}$ (first and second row) and velocities $v_{x/y}$ (third and forth row) for the different interpolation schemes (one per column) between two Wannier functions for monolayer $\mathrm{MoS}_2$ ($6\times 6\times1$ $\bk$-grid) of the CB-VB Wannierization.
    The corresponding matrix elements of the interpolated ab-initio velocities $v^{\mathrm{abi}}_x$ and $v^{\mathrm{abi}}_y$ are depicted in row three and four on the right side, respectively.
    The iso-surfaces of the two Wannier functions are displayed in b) and c), which are conduction and valence band only Wannierized, respectively.
    d) Location of the rhombic BZ used in a).
    }
    \label{fig:rpv}
\end{figure}

The accuracy of the interpolated Berry connection is not directly accessible because of its non-locality in $\bk$.
Instead, we determine the accuracy of the velocity operator as a proxy.
On the one hand, the scheme-dependent velocity $\bv^{\mathrm{S,\bk}}$ is computed from the Berry-connection $A^{\mathrm{S},\bk}$ using Eq.~\eqref{eq:velocity} for all schemes $\mathrm{S}$. 
On the other hand, the velocity $\bv^{\mathrm{abi},\bk}$ is obtained during the ab-initio calculation for each $\bk$-point independently and afterwards interpolated to arbitrary crystal momenta analogous to the other matrix elements.
In particular it is based on the same semi-unitary transformation matrices in Eq.~\eqref{eq:BlochToWan} that are used to obtain the Hamiltonian matrix elements in Wannier basis.
We start with a qualitative comparison of $\bv^{\mathrm{abi},\bk}$ with $\bv^{\mathrm{S,\bk}}$ for the different schemes.
Figure~\ref{fig:rpv} depicts the BZ-resolved Berry connection for the different interpolation schemes and the corresponding velocity matrix elements between two selected orbitals of the CB-VB Wannierization with $N_\bk=6$ for $\mathrm{MoS}_2$.
The Berry connection and velocity matrix elements show similar features for all interpolation schemes, except for the MV-scheme.
The MV-scheme shows (in this example) a reduced magnitude of all the matrix elements compared to the other schemes.
Examining the sym-, Lihm-, log-, and sclog-schemes in detail, we see differences in the maximum of the computed Berry connection.
For instance, the sclog-scheme predicts the largest Berry connection $|A_x|$ (in $x$-direction) at the $\Gamma$-point and also features the overall largest $|A_y|$.
These differences translate to the velocity operator as well: The magnitude of the $v_x$ and $v_y$ components are increased at the same positions of the BZ.
The ab-initio computed velocity operator, shows enhancements at the same locations.

\begin{figure*}[tb]
    \centering
    \includegraphics[width=\textwidth]{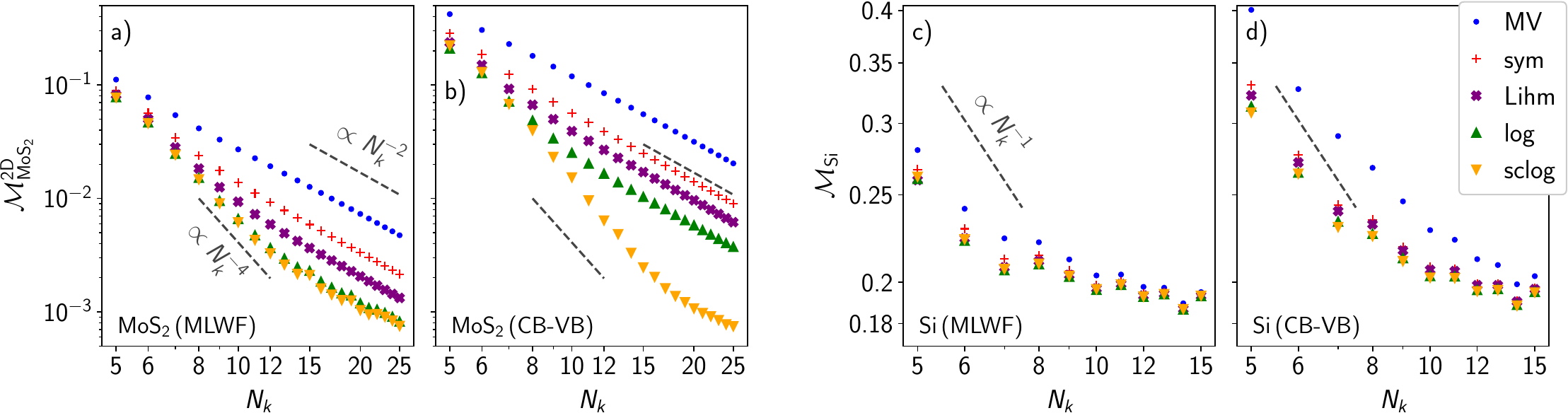}
    \caption{Mismatch $\mathcal{M}$, see Eq.~\eqref{eq:mismatch}, of the velocity operator for different schemes and ab-initio grid sizes calculated for
             $\mathrm{MoS_2}$: a) MLWF,
                               b) CB-VB,
             $\mathrm{Si}$: c) MLWF,
             and d) CB-VB.
             The dashed lines are guide for the eyes and indicate power scalings.
    }
    \label{fig:pva}
\end{figure*}

To compare the interpolation schemes quantitatively, one must account for all matrix elements of the velocity operator.
Therefore, we define a BZ-averaged mismatch measure for an interpolation scheme
\begin{align} \label{eq:mismatch}
    \mathcal{M}^{\mathrm{S}} = \sqrt{\frac
                         {\int\limits_{\mathrm{BZ}} \Vert \bv^{\mathrm{S,\bk}} - \bv^{\mathrm{abi},\bk}\Vert^2 \mathrm{d}\bk}
                        { \int\limits_{\mathrm{BZ}}\Vert\bv^{\mathrm{abi},\bk}\Vert^2 \mathrm{d}\bk}
                    }
,
\end{align}
where $\mathrm{S}$ is either MV, sym, Lihm, log, or sclog and $\bv^{\mathrm{S,\bk}}$ is computed according to Eq.~\eqref{eq:velocity}.
This measure is gauge independent as the Frobenius norm is invariant under unitary transformations and is directly comparable between different ab-initio grid sizes.
We note that one can also define a $\bk$-resolved mismatch for a more detailed investigation of specific systems, but this is beyond the scope of this study.
Figure~\ref{fig:pva} shows this mismatch for the interpolation schemes as a function of the ab-initio grid size $N_k$ in double-logarithmic plots for the MLWF and CB-VB Wannierizations of $\mathrm{MoS}_2$ and $\mathrm{Si}$.
The BZ integrals inside the mismatch definition, see Eq.~\eqref{eq:mismatch}, where evaluated on a $\bk$-grid twice as large as the Wigner-Seitz cell interpolation grid to completely cover the commutator during the evaluation of the velocity operator defined in Eq.~\eqref{eq:velocity} \footnote{Due to the unitarity of the Fourier transform, the integrals can be equivalently evaluated in real-space, where the commutator is translated into a convolution with a maximum support (in a mathematical sense) of twice the Wigner-Seitz cell in each direction.}.

We first discuss the behavior for $\mathrm{MoS}_2$, which has a very small basis spill due to its energetically separated bands, see Sec.~\ref{sec:basisSpillQuantification} for details.
The MV-, sym-, and Lihm-scheme have a mismatch proportional to $N_k^{-2}$, i.e., $\mathcal{O}(|\bb|^2)$ for the MLWF case (panel a) as well as for the CB-VB case (panel b), which is in agreement with the derived error scalings.
Their relative ordering can be understood from analytical observations:
The sym-scheme improves the MV-scheme, because the respective leading error coefficients (in $\bb$) are smaller, which follows from a Taylor expansion of the $M^{\bk\bb}$ matrices in Eqs.~\eqref{eq:sym} and \eqref{eq:MV}, respectively.
The Lihm-scheme yields a lower mismatch, as its $\bk$-independent basis change, see discussion in Sec.~\ref{sec:Lihm}, leads to effective overlap matrices, which are on average closer to unity than those of the raw sym-scheme, resulting in smaller expansion errors.
The log- and sclog-scheme are for both Wannierzations consistently better than the Lihm-scheme as they account for the matrix structure properly.
In the MLWF case, the sclog-scheme is only marginally better than the log-scheme, because of the very slow change of the overlap matrices (due to their maximal localization in the supercell).
In the CB-VB case, only the sclog-scheme has a better than quadratic error scaling (for $N_k \leq 18$), which is a clear indication of the effectiveness of the recursive refinement.
Note that the sclog-scheme is the only one that reaches the same mismatch as for the MLWF case for $N_k \geq 18$.
For these grid sizes the mismatch scales with $\mathcal{O}(|\bb|^2)$, which is the inevitable error scaling caused by the disagreement between $\log M^{\bk\bb}$ and $\log M^{\mathrm{complete},\bk\bb}$.

The effect of the basis spill is much more pronounced for $\mathrm{Si}$, where no power law of the error scaling arises, see panels c and d. 
In both cases, the MV-scheme results in the largest mismatch, followed by the sym-, and Lihm-schemes.
The log- and sclog-schemes are nearly indistinguishable and provide the lowest mismatch.
As the matrix logarithm is a rough approximation in this case, the log- and sclog-schemes improve the Lihm-scheme only slightly.
Similarly to $\mathrm{MoS}_2$, the MLWF Wannierzations yield a lower mismatch than the CB-VB ones.
In Appendix~\ref{app:Mcut}, we define an energy truncated mismatch restricted to the frozen energy window used during the disentanglement procedure.
Applying it to $\mathrm{Si}$ results in a similar behavior of the interpolations schemes as the mismatch for $\mathrm{MoS}_2$, in particular a power-law like scalings are observed.

\subsection{Quantification of basis incompleteness} \label{sec:basisSpillQuantification}
\begin{figure}[tb]
    \centering
    \includegraphics[width=\columnwidth]{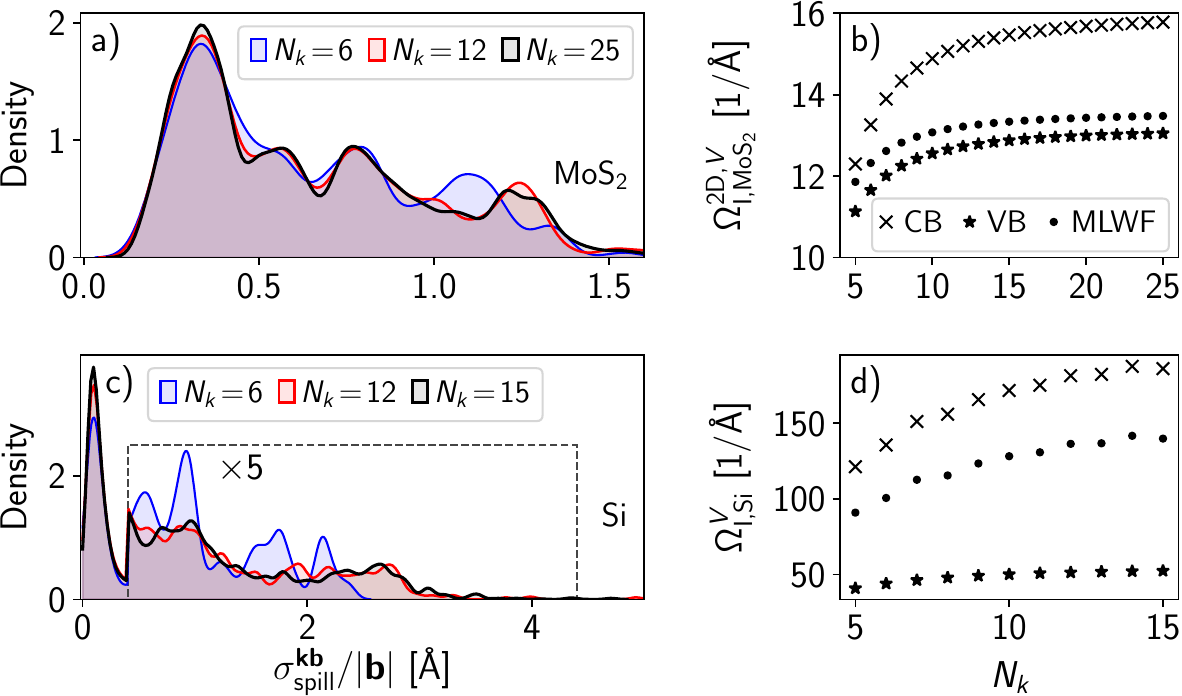}
    \caption{Quantification of basis spill of the MLWF Wannierizations for different ab-initio grid sizes for a), b) $\mathrm{MoS}_2$ (only in-plane components) and c), d) $\mathrm{Si}$.
             a) and c) kernel density estimation (Gaussian width 0.1\,\r{A}) of the singular values of the infinitesimal spill.
             b) and d) Invariant part of the Wannier function spread as function of the ab-initio grid size for only conduction bands (CB, black cross), conduction and valence bands (MLWF, big black dot), and valence bands only (VB, black star).
             Note that d) is coincides with the CB-VB Wannierization as no disentanglement for $\mathrm{MoS}_2$ was performed.
    }
    \label{fig:basisSpill}
\end{figure}

The basis set incompleteness discussed in Sec.~\ref{sec:basisSpill} is the main cause of the sclog-scheme yielding a systematic error in the calculation of the velocity operator.
In Fig.~\ref{fig:basisSpill} we quantify the basis incompleteness for $\mathrm{MoS}_2$ in panels a and b, and for $\mathrm{Si}$ in panels c and d.
The kernel density estimations in Fig.~\ref{fig:basisSpill}a and c show that in both cases the distribution of the infinitesimal spill singular values $\sigma^{\bk\mathbf{e}_\bk}_{\mathrm{spill}}$ is almost converged for $N_\bk=12$ for $\mathrm{MoS}_2$ as well as for $\mathrm{Si}$, when comparing it to the finer $\bk$-grids.
This reflects, that the infinitesimal spill $\sigma^{\bk\bb}_n/|\bb|$ converges for reasonable ab-initio grid sizes and indicates that Eq.~\eqref{eq:infinitesimalSpill} is applicable for these grid sizes.
However, the invariant part of the Wannier function spread $\Omega^V_{\mathrm{I}}$, see panels b and d, is not converged for $N_\bk=12$ and converges only slowly with respect to the grid size.
In absolute values, the basis spill for $\mathrm{MoS}_2$ is much smaller than that of $\mathrm{Si}$.
The $\Omega^V_{\mathrm{I}}$ for the different subspaces converge faster whenever no disentanglement is required, i.e., for the valence band-only subspaces and for $\mathrm{MoS}_2$ (as no disentanglement of the conduction bands was performed).
This indicates that the disentanglement itself is the reason for this slow convergence.
The converged value of $\Omega_{\mathrm{I}}^V$ is directly linked to the error in the estimated Berry connection, see Eq.~\eqref{eq:infinitesimalSpill} and provides a limit for the accuracy of the estimated Berry connection.
The much larger value of $\Omega^V_{\mathrm{I}}$ for $\mathrm{Si}$ compared to $\mathrm{MoS}_2$ is in alignment with the larger mismatch for $\mathrm{Si}$.

\subsection{Optical conductivity}

\begin{figure*}[tb]
    \centering
    \includegraphics[width=\textwidth]{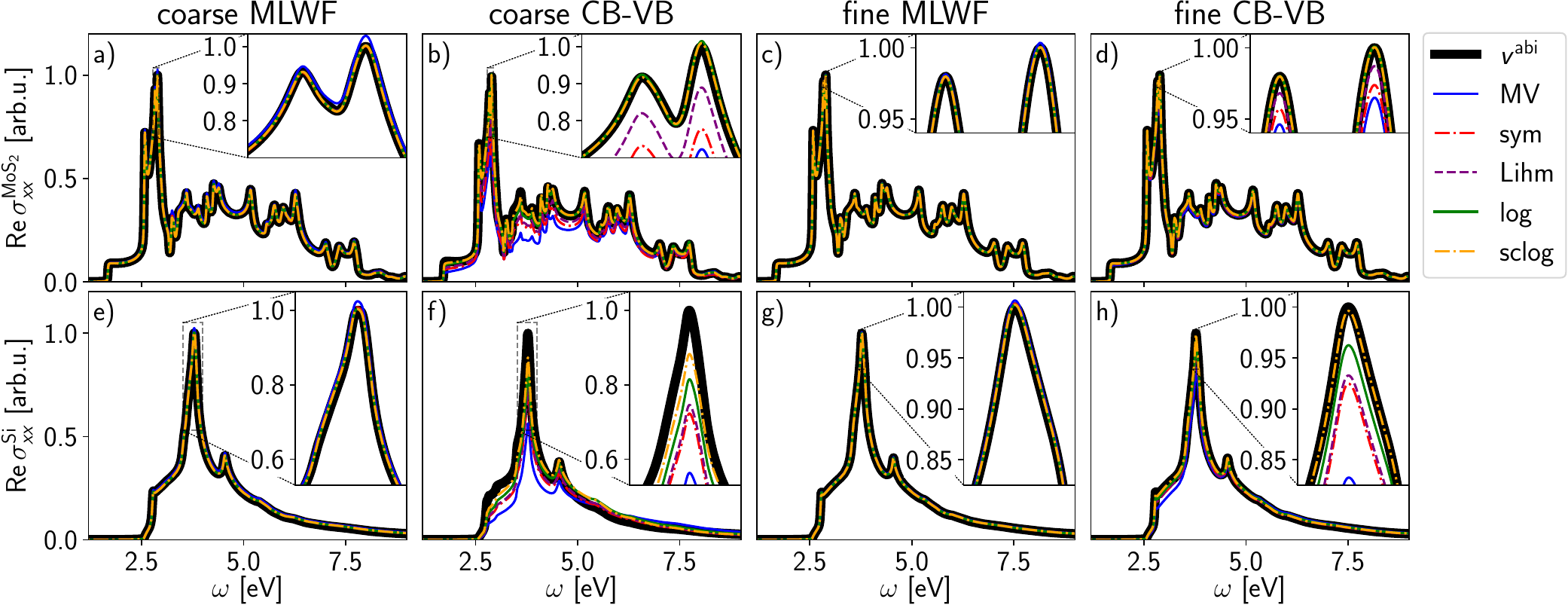}
    \caption{Real part of the optical conductivity for $\mathrm{MoS}_2$ (a-d) and $\mathrm{Si}$ (e-h) obtained using the velocity matrix elements of different Berry connection interpolations schemes (colored lines) and Wannierizations.
    The spectra obtained from the ab-initio computed velocity matrix elements (thick black lines) are normalized.
    $\mathrm{MoS}_2$: a) $N_k=8$ MLWF, b) $N_k=8$ CB-VB, c) $N_k=25$ MLWF, d) $N_k=25$ CB-VB.
    $\mathrm{Si}$: e) $N_k=8$ MLWF, f) $N_k=8$ CB-VB, g) $N_k=15$ MLWF, h) $N_k=15$ CB-VB.
    }
    \label{fig:absorptionSpectra}
\end{figure*}
\begin{figure}[tb]
    \centering
    \includegraphics[width=0.99\columnwidth]{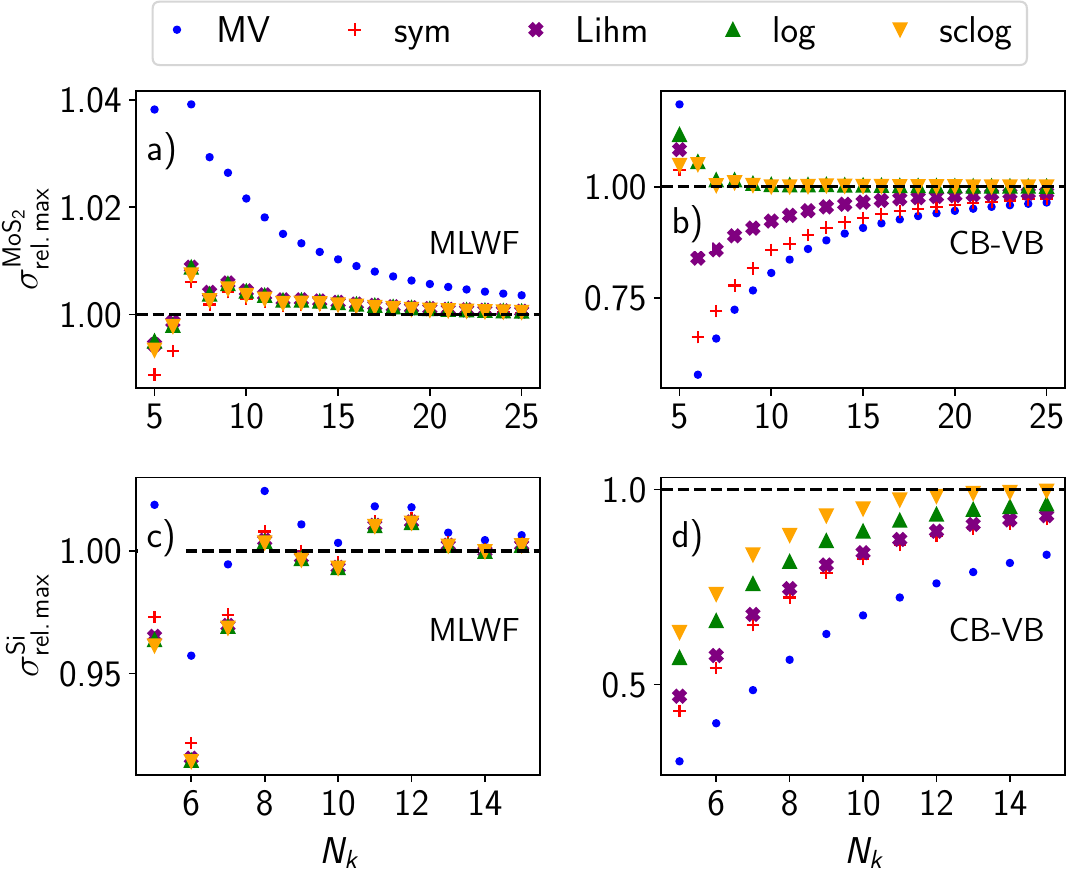}
    \caption{Scheme-dependent relative maximal peak height of the real part of the optical conductivity for different grid sizes and Wannierizations compared to the one computed from the ab-initio velocity operator.
     $\mathrm{MoS}_2$: a) MLWF,
                        b) CB-VB, 
    $\mathrm{Si}$: c)  MLWF, 
        and         d)  CB-VB. 
    }
    \label{fig:maxAbsorption}
\end{figure}

To demonstrate the influence of the Berry connection calculated from the different schemes on a measurable physical quantity, we calculate the real part of the optical conductivity. 
Up to a system dependent factor, the optical conductivity at zero temperature is given by \cite{greenwood_1958}
\begin{align}
\sigma_{\!\alpha\beta}(\omega) \propto \mathrm{i}\sum_{c, v} \intBZdk\,\frac{(\bv^\bk_{vc})_{\alpha} (\bv^\bk_{vc})^{*}_{\beta}}{(E^\bk_v - E^{\bk}_c)(\omega + E^\bk_v - E^\bk_c - \mathrm{i}\eta)}
,
\end{align}
where the indices $c$ and $v$ run over the valence and conduction bands with corresponding energies $E^{\bk}_{c/v}$, the asterisk indicates complex conjugation, and $\alpha$, $\beta$ denote arbitrary real-space directions.
In our numerical evaluation, we use $\eta = 0.1\,\mathrm{eV}$ and evaluate the integral for $N_k=500$. 
Figure~\ref{fig:absorptionSpectra} shows the real part of the optical conductivity for monolayer $\mathrm{MoS}_2$ (panels a-d) in zig-zag direction and for bulk $\mathrm{Si}$ (panels e-h) in an arbitrary direction ($\alpha=\beta=x$).
Among the displayed spectra, the ones based on the ab-initio calculated and Wannier interpolated velocities $v^{\mathrm{abi},\bk}$ (black lines) serve as reference.
For the MLWF Wannierizations, all the schemes match the reference very closely.
However, for the CB-VB Wannierizations there are strong differences between the schemes, with the sclog performing best and the MV worst.
The optical conductivity computed for $\mathrm{Si}$ in the CB-VB Wannierization is only reproduced to very good agreement by the sclog-scheme for the fine-grid.
We quantify the deviations for the optical conductivity via the maximal peak height for each scheme $S$ and Wannierization by
\begin{align}
    \sigma_{\mathrm{rel.\,max}} = \frac{\max\Re\sigma_{S,xx}}{\max\Re\sigma_{\mathrm{abi},xx}},
\end{align}
which we present in Fig.~\ref{fig:maxAbsorption} as function of $N_k$.
We asserted that the relative peak height approaches the limit of 1 at least quadratically for all the schemes and both Wannierization types.
In the MLWF case, only the MV-scheme differs significantly from the others (most times worse).
The other schemes show relative errors well below $1\%$ ($10\%$) for $\mathrm{MoS}_2$ ($\mathrm{Si}$).
Contrary, in the CB-VB case, the relative errors are very different for the schemes and reach up to $28\%$ ($44\%$) for $\mathrm{MoS}_2$ ($\mathrm{Si}$) for $N_k=8$.
The sclog-scheme features in this case for $\mathrm{MoS}_2$ an error of only $0.1\%$, thus being practically independent of the Wannierization.
Due to the high basis spill of the $\mathrm{Si}$ Wannierization, the sclog-scheme still has a significant error, which goes below $1\%$ for $N_k=13$, where the MV-scheme has an error of $21\%$.

The deviation of the maximum conductivity is much stronger scheme-dependent than the mismatch discussed in Sec.~\ref{sec:comparisonVM} because the optical conductivity weights the velocity operator stronger at energies closer to the Fermi level.
We emphasize that only the sclog-scheme allows for a nearly Wannierization-independent computation of absorption spectra for reasonable ab-initio grid sizes.

\section{Summary and outlook} \label{sec:sao}
We assessed the quality of existing and newly developed interpolation schemes, i.e., the Marzari-Vanderbilt, the symmetric, the Lihm, the logarithmic, and the self-consistent logarithmic scheme, for the Berry connection analytically as well as numerically.
We proposed a logarithmic interpolation scheme based on the notion that the Berry connection is a generator of the overlap matrices $M^{\bk\bb}$, in contrast to previous studies, which compute the Berry connection element-wise.
We extended it to a self-consistent logarithmic scheme via recursive refinement based on the explicit evaluation of the path-ordered matrix exponential by employing a Magnus expansion.
To compare the different interpolation schemes, we introduced a mismatch measure between the ab-initio interpolated and scheme-dependent velocity operator.
The self-consistent logarithmic interpolation scheme features the fastest reduction of the mismatch with respect to the ab-initio grid size for monolayer $\mathrm{MoS}_2$ for two different Wannierizations.
For $\mathrm{Si}$, all interpolation schemes result in a much higher mismatch with the logarithmic scheme and its self-consistent extension was consistently performing best.
We identified the invariant part of the Wannier-function spread $\Omega_\mathrm{I}^V$ as a fundamental limitation of the accuracy.
Lastly, we highlighted that the proposed self-consistent logarithmic interpolation scheme enables faster convergence of optical conductivities with respect to the ab-initio grid size by comparing it to the ones based on the ab-initio velocity operator.
For $\mathrm{MoS}_2$ ($\mathrm{Si}$) the deviation of the maximal optical conductivity is reduced from $3.5\%$ ($17\%$) for the Marzari-Vanderbilt-scheme to $0.04\%$ ($0.2\%$) for $N_k=25$ ($N_k=15)$ for the stitched Wannierization of conduction and valence bands.
We expect similar effects for other often used Wannier gauges, like the ones that enforce the symmetry constraints of the crystal explicitly \cite{sakuma_symmetry-adapted_2013,koepernik_symmetry-conserving_2023,koretsune_construction_2023}.

As we cannot control the basis spill after the disentanglement and the Wannierization, we propose to evaluate the symmetric, Lihm, logarithmic and consistent logarithmic scheme in numerical calculations and to automatically select the one yielding the lowest mismatch compared to the ab-initio velocity operator.
Because of the recursive refinement of the self-consistent logarithmic scheme and the inevitable error introduced by the basis spill, we do not expect that higher-order interpolation schemes based on more overlap matrices $M^{\bk\bb}$ will lead to more accurate results.
In a subsequent study, it should be investigated by the means of the ab-initio velocity operator under which circumstances the recently proposed schemes \cite{lihm2026accurate,cole_exact_2026} can be combined with our scheme to further improve the quality of the estimated Berry connection.
An interesting question to investigate next is whether the self-consistent logarithmic interpolation scheme preserves the symmetries of the crystal.
Additionally, a band energy-weighted mismatch measure of the velocity operator may enable a more quantitative estimation of deviations for energy-dependent observables like the optical conductivity.
We provide our reference implementation in \cite{thuemmler_2026}.
Lastly, we emphasize that the methodology of the self-consistent logarithmic scheme can be transferred to composite operators, to compute quantities like the spin Hall conductivities \cite{ryoo_computation_2019} as well.

\section*{Acknowledgment}
The project was funded by the Deutsche Forschungsgemeinschaft (DFG, German Research Foundation) – Project-ID 398816777 – SFB 1375, A1. 
We thank J.-M. Lihm for his comments on the initial version of this manuscript.

\appendix

\section{Details of Fourier interpolation} \label{app:FTdetail}
In the Fourier transforms used in this manuscript, we omit the interpolation details for the sake of notation.
To interpolate any quantity $Q^{\bk}$ in momentum space, we first compute its discrete Fourier transform
\begin{align}
    Q^\bR_{\mathrm{grid}} = \frac{1}{N_\bk}\sum_\bk \expi{\bk\bR} Q^\bk
\end{align}
on a uniform grid.
This quantity is then translated to the Wigner-Seitz (WS) cell as
\begin{align}
   Q^{\bR}_{\mathrm{WS}} = \frac{ Q^{\mathrm{reduce}(\bR) } } {n^\bR_{\mathrm{degen}}}
,
\end{align}
where the $\mathrm{reduce}$ function maps the WS lattice vector $\bR$ to the supercell lattice vector and $n^\bR_{\mathrm{degen}}$ is the number of WS lattice vectors originating from the same supercell lattice vector.
Any additional phase factors in the Fourier transforms, e.g. $\expi{\frac{\bb\bR}{2}}$, are applied to $Q^{\bR}_{\mathrm{WS}}$.
The interpolation back to momentum space is done exactly as given by Eq.~\eqref{eq:opRtoK}.
This Fourier interpolation scheme is the same as in Wannier90 \cite{pizzi_2020} with the option $\mathrm{use\_ws\_distance}=0$.

\section{Proof of $(M^{\bk-\frac{\bb}{2}, \bb})^\dagger  =  M^{\bk+\frac{\bb}{2}, -\bb}$} \label{app:Mdagger}
To prove Eq.~\eqref{eq:MdaggerSym}, we start from Eq.~\eqref{eq:defMR} to show that
\begin{align} \label{eq:MRdagger}
    (M^{\bR\bb})^\dagger &= \frac{1}{N_\bk} \sum\limits_{\bk} \exmi{\bk\bR} M^{\bk+\bb, -\bb} \nonumber \\
    &= \frac{1}{N_\bk}\sum\limits_{\bk} \exmi{(\bk-\bb)\bR} M^{\bk, -\bb} \\
    &= \expi{\bb\bR}  M_{-\bR,-\bb} \nonumber   
,
\end{align}
where we used $(M^{\bk,\bb})^\dagger = M^{\bk+\bb, -\bb}$.
Plugging Eq.~\eqref{eq:MRdagger} into the inverse Fourier transform of Eq.~\eqref{eq:defMR} yields
\begin{align}
    (M^{\bk-\frac{\bb}{2}, \bb})^\dagger = \sum\limits_{\bk} \expi{(\bk+\frac{\bb}{2})\bR} M_{-\bR, -\bb} \expi{\bb\bR} = M^{\bk+\frac{\bb}{2}, -\bb}
.
\end{align}

\section{Branch selection using guiding Wannier centers} \label{app:branch}
The principal branch of the matrix logarithm in Eq.~\eqref{eq:log} is not always the physically sound one, especially if the out-of-plane component of monolayer materials or a $\Gamma$-point only sampling is considered.
To select the branch meaningfully, the guiding centers of the Wannier functions $\mathcal{R}_i$ are employed (similar to $\mathrm{use\_guiding\_centres=1}$ in Wannier90).
The matrix logarithm is evaluated using the Schur-Parlett algorithm \cite{davies2003schur} with special care taken for near degenerated eigenvalues.
First, the Schur-decomposition is computed
\begin{align}
    M^{\bk\bb} = U^{\bk\bb} T^{\bk\bb} U^{\bk\bb\dagger},
\end{align}
where $U^{\bk\bb}$ is unitary and $T^{\bk\bb}$ upper triangular.
Second, the matrix logarithm of $\log T^{\bk\bb}$ is evaluated for its diagonal elements as
\begin{align}
    (\log T^{\bk\bb})_{ii} = \ln T^{\bk\bb}_{ii} + 2\pi \mathrm{i} K^{\bk\bb}_i,
\end{align}
where $\ln$ denotes the principal branch of the natural logarithm and $K^{\bk\bb}_i$ are integer numbers chosen to minimize
\begin{align}
    \Vert \log M^{\bk\bb} - \mathrm{i} \bb\mathcal{R}\Vert = \Vert \log T^{\bk\bb} - \mathrm{i} \bb U^{\bk\bb,\dagger}\mathcal{R} U^{\bk\bb}\Vert ,
\end{align}
where $\mathcal{R}_{ij} = \delta_{ij} \mathcal{R}_i$.
In practice it suffices to minimize $|(\log T^{\bk\bb})_{ii} - \mathrm{i} (\br U^{\bk\bb,\dagger}\mathcal{R}U^{\bk\bb})_{ii}|$ for each index $i$ independently as the $T^{\bk\bb}$ matrices are diagonal dominated.
The off-diagonal elements of $\log T^{\bk\bb}$ are computed as in Ref.\,\cite{al-mohy_improved_2012}.
\begin{figure}[hptb]
    \centering
    \includegraphics[width=0.99\columnwidth]{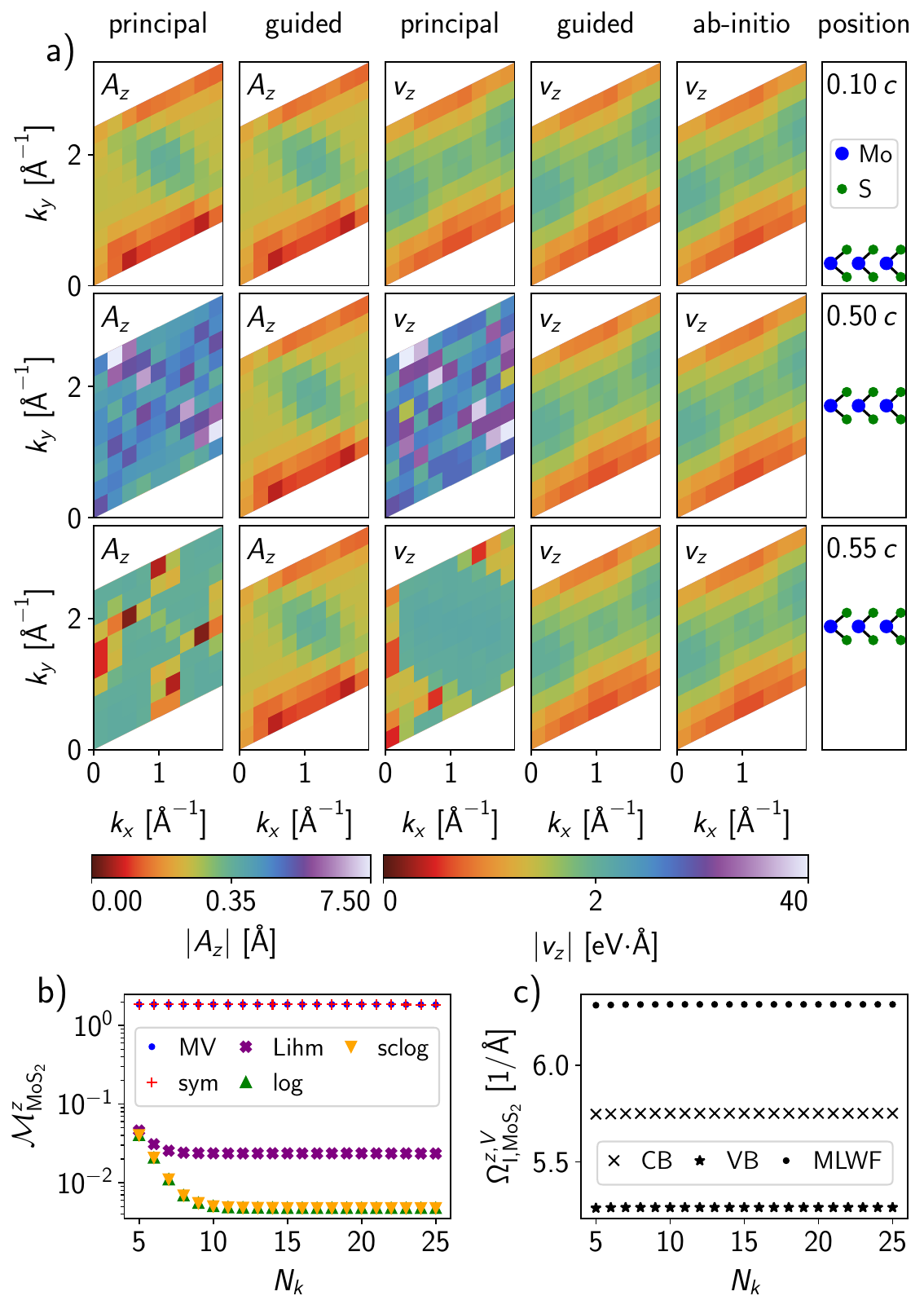}
    \caption{Effect of the branch selection method for the (sc)log-scheme in the $z$ (out-of-plane) direction of $\mathrm{MoS}_2$ monolayer for the same orbitals and configurations as in Fig.~\ref{fig:rpv}.
    a) Each row depicts the Berry connection and the velocity matrix (without interpolation) in Wannier basis for the geometry shown in a projection of a $3\times 3\times 1$ supercell on the right most axes.
    The $z$-component of $\mathrm{Mo}$ is written in terms of the $c$-lattice constant (=25\,\AA) in the supercell depiction.
    The color map values are linearly interpolated between the indicated values.
    b) Mismatch along the $z$-component as function of the ab-initio grid size for the second geometry computed with guiding centers.
    c) Invariant part of the Wannier function spread evaluated in $z$-direction only as function of the ab-initio grid- size.
    }
    \label{fig:zComponent}
\end{figure}
Figure~\ref{fig:zComponent}a shows the out-of-plane component of the computed Berry connection for $\mathrm{MoS}_2$ placed at different positions of the unit cell.
Selecting the branches of the logarithms in Eq.~\eqref{eq:log} by employing the guiding centers gives the same off-diagonal matrix elements in very good agreement to the corresponding ab-initio computed velocity matrix elements.
When employing the principal branch, the Berry connection exhibits large jumps between close $\bk$-points, impeding any Fourier interpolation.
The most severe errors arise, when the monolayer is centered in the middle of the unit cell as the Wannier centers $r_n$ correspond to $\pm \pi$ phases of $\exmi{\bb\br_n}$, which are at the border of the principal branch.
When employing the guiding centers, the mismatch is best for the (sc)log-scheme, see Fig.~\ref{fig:zComponent}b.
Note that the MV- and sym- as well as the log- and sclog-scheme are equivalent along that direction because of $N_{k,z} =1$.
Otherwise, the mismatch features the same ranking of the different interpolation schemes as for the in-plane components.
As only the in-plane components are refined with increasing $N_k$ the mismatch converges fast, but not as rapid as the $\Omega^{z,V}_{\mathrm{I}}$ shown in Fig.~\ref{fig:zComponent}c. 

\section{Alternative logarithmic interpolation scheme} \label{app:altLogScheme}
According to Eq.~\eqref{eq:logM}, we define an alternative logarithmic interpolation scheme as
\begin{align} \label{eq:altlog}
    \bD^{\mathrm{alt},\bk} = \mathrm{i} \sum\limits_{\bb} w_{\bb} \bb \log M^{\bk-\frac{\bb}{2},\bb}
        .
\end{align}
The Hermiticity of $\bD^{\mathrm{alt},\bk}$ follows directly from Eqs.~\eqref{eq:MdaggerSym} and \eqref{eq:weightCondition}.
The scheme is independent of the origin as a shift by $\mathcal{R}$, see Eq.~\eqref{eq:BerryTI}, leads to $M^{\bk\bb} \rightarrow M^{\bk\bb} \exmi{\bb\mathcal{R}}$, and thus, as this phase factor is constant with respect to $\bk$, to 
\begin{align}
    \log M^{\bk-\frac{\bb}{2},\bb} \rightarrow \log M^{\bk-\frac{\bb}{2},\bb} -\mathrm{i}\bb\mathcal{R}
    ,
\end{align}
which results in $\bD^{\mathrm{alt},\bk} \rightarrow \bD^{\mathrm{alt},\bk} + \mathcal{R}$ in accordance with Eq.~\eqref{eq:BerryTrafo}.
Finally we note that one could directly compute the $M^{\bk-\frac{\bb}{2},\bb}$ matrices from an ab-initio calculation with doubled grid size in each direction to combine the benefits of the $\bD^{\mathrm{log},\bk}$ and the $\bD^{\mathrm{alt},\bk}$ scheme.

\section{Comments on size-consistency} \label{app:sizeConsistent}
The log- and sclog- scheme are size-consistent (SC) with respect to the polarization, meaning that the same sum of the Wannier centers is obtained from a BZ sampling with $N_k\times N_k\times N_k$ points and a supercell consisting of $N_k\times N_k\times N_k$ cells that is only sampled at the $\Gamma$-point \cite{stengel_accurate_2006,lihm2026accurate}.
To show that, we consider the Wannier functions
\begin{align}
 \ket{u^{\mathrm{SC}}_I} = \ket{\bR_I i}
\end{align}
described by the index $I=(i, R_I)$ with the lattice vector $R_I$ as the cell periodic functions of the supercell.
The supercell overlap matrix elements are therefore, see Appendix~A of Ref.\cite{lihm2026accurate}, given by
\begin{align}
M^{\mathrm{SC},\bb}_{IJ} &= \frac{\exmi{\bb\bR_J}}{N_\bk} \sum\limits_{\bk} M^{\bk\bb}_{ij} \expi{\bk(\bR_I - \bR_J)}.
\end{align}
To evaluate the trace of the matrix logarithm, we define $\tilde{M}^{\mathrm{sc},\bb}_{IJ} = M^{\mathrm{sc},\bb}_{IJ} \expi{\bb\bR_J}$ and note that it straight-forward to show by induction that for $n \geq 1$
\begin{align}
  \left[\left(\tilde{M}^{\mathrm{SC},\bb}\right)^n\right]_{IJ} = \frac{1}{N_\bk}\sum\limits_{\bk} \left[\left(M^{\bk\bb}\right)^n\right]_{ij} \expi{ \bk(\bR_I - \bR_J)}
\end{align}
holds, which allows to conclude
\begin{align} \label{eq:logMkSum}
    \left(\log \tilde{M}^{\mathrm{SC},\bb}\right)_{IJ} = \frac{1}{N_\bk} \sum\limits_{\bk} \left(\log M^{\bk\bb}\right)_{ij} \expi{ \bk(\bR_I - \bR_J)}
\end{align}
using an intermediate formal power series representation of the matrix logarithm.
By employing $\Tr\log (AB) = \Tr \log A+ \Tr\log B$ and Eq.~\eqref{eq:logMkSum}, we obtain
\begin{align}
    \Tr \log M^{\mathrm{SC},\bb} &= \Tr\log(\exmi{\bb\,\mathrm{diag}(\bR_I)}) + \frac{1}{N_\bk}\Tr \log \sum\limits_{\bk} M^{\bk\bb} \nonumber \\
                                 &= -\mathrm{i}\bb \sum_I \bR_I + \frac{1}{N_\bk}\Tr\sum\limits_{\bk} \log M^{\bk\bb}
\end{align}
and thus
\begin{align} \label{eq:polSize}
   \Tr \bD^{\mathrm{SC}} &= \mathrm{i} \sum\limits_{\bb} w_\bb \Tr \log M^{\mathrm{SC},\bb} \nonumber \\
                &= \sum\limits_{I} \bR_I + \frac{\mathrm{i}}{N_\bk}\Tr\sum\limits_{\bb\bk} w_\bb\log M^{\bk\bb} \nonumber \\
                &= \sum\limits_{I} \bR_I + \Tr \bD^{\bR}
,
\end{align}
which proves the polarization size-consistency as the sum over the lattice vectors $\bR_I$ account for the unit cell shifts of the polarization that map the unit cells to the supercell.
We remark, that the polarization obtained form the logarithmic scheme is identical to the Berry phase approach of Refs. \cite{king-smith_theory_1993,stengel_accurate_2006}.
The self-consistent scheme is polarization size-consistent as well, as the sum in the Magnus expansion in Eq.~\eqref{eq:MagnusExpansionInterpol} yields 
the same trace when integrated over $\bk$ as the trace of the commutator vanishes.
The polarization size-consistency of the alternative scheme in Appendix~\ref{app:altLogScheme} can be proven similarly.
The log- and sclog-scheme are not size-consistent in the sense that all the matrix elements of the Berry connection can be directly translated from the unit cell to the $\Gamma$-point only sampled supercell.
An element-wise agreement is not expected as the vector spaces spanned by the overlap matrices, i.e., small ones for each $\bk$-point compared to a single large one, are conceptually different.

\section{Convergence of the sclog-scheme} \label{app:scConvergence}
We investigated two types of iteration schemes to obtain the self-consistency of the logarithmic scheme.
When performing the fix-point iteration using the estimate of the Berry connection directly instead of relying on the $S^{\bk}_{\bb}$ matrices, the estimate for the logarithm of the path-ordered integral in Eq.~\eqref{eq:MagnusExpansionInterpol} is always anti-Hermitian, in contrast to $\log M^{\bk\bb}$ in an incomplete basis.
\begin{figure}[hptb]
    \centering
    \includegraphics[width=0.99\columnwidth]{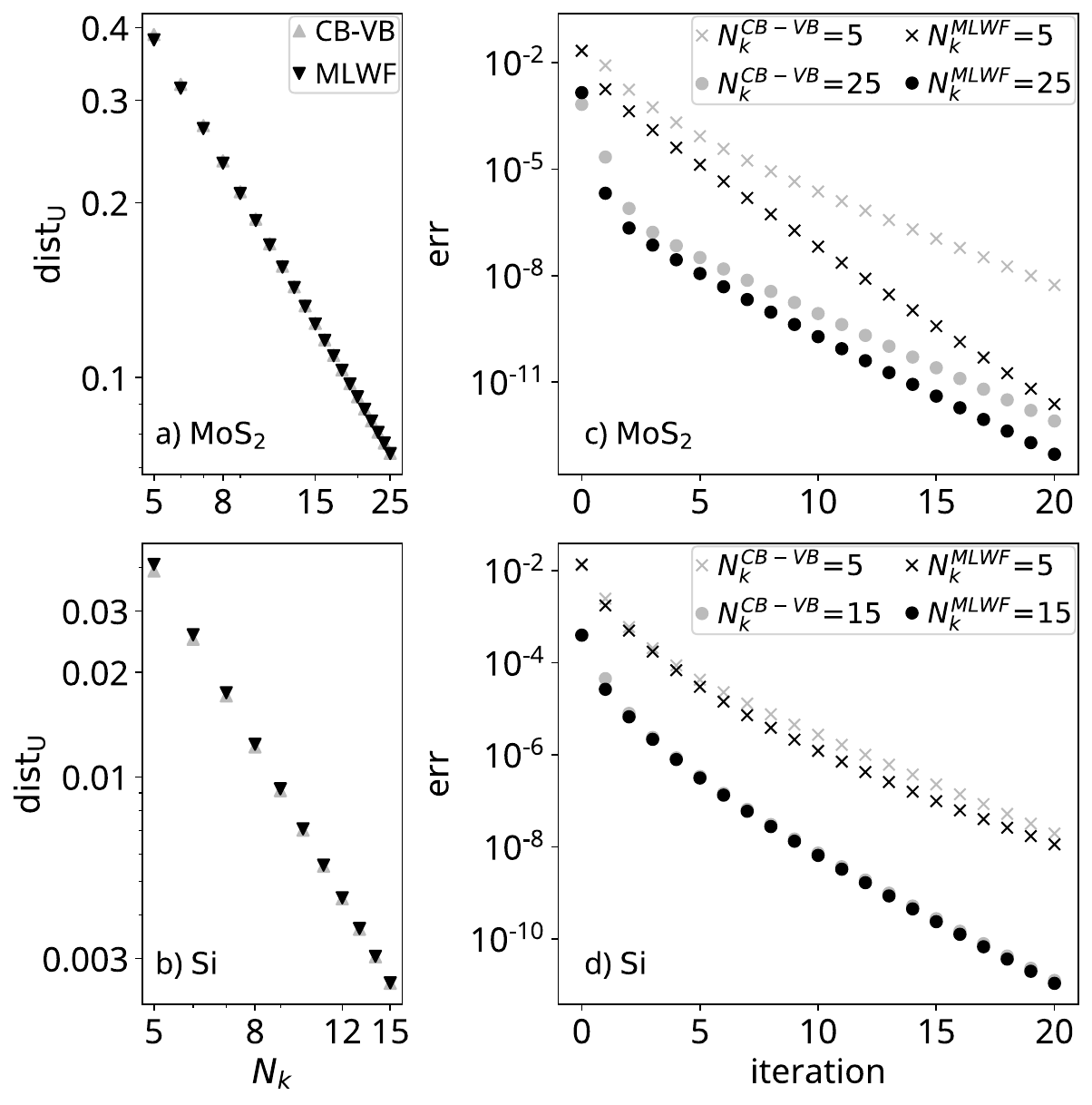}
    \caption{Norm of the Hermitian part of the matrix logarithm of the overlap matrices for a) $\mathrm{Si}$ and b) $\mathrm{MoS}_2$.
    Exemplary evolutions of the error of the fixed-point iteration of the sclog-scheme as function of the iteration count for c) $\mathrm{MoS}_2$ and d) $\mathrm{Si}$.}
    \label{fig:conv}
\end{figure}
The minimal distance between those is therefore $\Vert \log M^{\bk\bb} + \left(\log M^{\bk\bb}\right)^\dagger \Vert/2$.
In Fig.~\ref{fig:conv}a and b, we depict the averaged minimum distance
\begin{align}
   \mathrm{dist}_\mathrm{U} = \frac{1}{2N_\bk N_{\bb}} \sqrt{ \sum\limits_{\bk\bb} \Vert \log M^{\bk\bb} + \left(\log M^{\bk\bb}\right)^\dagger \Vert^2 } 
,
\end{align}
where $N_\bb$ denotes the number of $\bb$-vectors as function of the ab-initio grid size.
As shown in Fig.~\ref{fig:conv}c and d, this quantity is orders of magnitudes larger than the errors 
\begin{align}
    \mathrm{err}_n = \frac{1}{N_\bk N_\bb}\sqrt{\sum_{\bk\bb} \Vert I^{\bk\bb}_n - \log M^{\bk\bb}\Vert^2}
\end{align}
of the well converging fix-point iteration based on the non anti-Hermitian $S^{\bk}_{\bb}$ matrices in Eq.~\eqref{eq:MagnusExpansionInterpol}.

\section{Details of basis spill quantification} \label{app:basisSpill}
First, we prove Eq.~\eqref{eq:sigmaSpill} without relying on the sine-cosine decomposition.
As $M^{\mathrm{complete},\bk\bb}$ is unitary, we find
\begin{align} \label{eq:MkfullSum}
    1 &= |M^{\bk\bb}\mathbf{x}|^2 + |M^{\mathrm{spill},\bk\bb} \mathbf{x}|^2
\end{align}
for an arbitrary unit vector $\mathbf{x}$ of dimension $n_{\mathrm{W}}$.
Employing the min-max theorem for singular values \cite{horn2012matrix} to Eq.~\eqref{eq:MkfullSum} leads to Eq.~\eqref{eq:sigmaSpill}.
From Eq.~\eqref{eq:MkComplete} we derive
\begin{align}
    \frac{M^{\mathrm{spill},\bk\bb}}{\mathrm{i}|\bb|} = \mathbf{e}_\bb\bD^{\mathrm{spill},\bk} + \mathcal{O}(|\bb|)
\end{align}
to obtain the singular values of $\bD^{\mathrm{spill},\bk}\mathbf{e}_\bb$ under a linear perturbation in $|\bb|$ \cite{stewart1998perturbation}
\begin{align} \label{eq:sigmaSpillUnit}
    \sigma^{\mathrm{spill},\bk\mathbf{e}_\bb}_n = \frac{\sigma^{\mathrm{spill},\bk\bb}_n}{|\bb|} + \mathcal{O}(|\bb|)
.
\end{align}
In the limit $\bb \rightarrow \bZero$, the infinitesimal spill is therefore bounded as given in Eq.~\eqref{eq:infinitesimalSpill}.
The singular values of $M^{\mathrm{spill},\bk\bb}$ are connected to the invariant part of the Wannier function spread $\Omega_{\mathrm{I}}$ \cite{marzari_1997} via
\begin{align} \label{eq:OmegaI}
    \Omega_{\mathrm{I}} &= \frac{1}{N_\bk} \sum_{\bk\bb} w_{\bb} \sum_{m=1}^{n_\mathrm{W}} \left[ 1 - \sum_{n=1}^{n_\mathrm{W}} |M^{\bk\bb}_{mn}|^2 \right] \nonumber \\
    &= \frac{1}{N_\bk} \sum_{\bk\bb} w_{\bb} \sum_{n=1}^{n_\mathrm{W}} (\sigma^{\mathrm{spill},\bk\bb}_n)^2
.
\end{align}
By replacing the sums in Eq.~\eqref{eq:OmegaI} with a BZ integral over the unit cell volume $V$
\begin{align}
    \frac{1}{N_\bk}\sum_{\bk} \rightarrow \frac{V}{(2\pi)^3}\intBZdk
    ,
\end{align}
the resulting quantity
\begin{align}
 \Omega_{\mathrm{I}}^V \equiv \frac{(2\pi)^3}{V} \Omega_{\mathrm{I}} &=  \sum\limits_{\mathbf{e}_{\bb}} w_{\mathbf{e}_{\bb}} \sum\limits_{n=1}^{n_{\mathrm{W}} } \,\intBZdk  \left(\frac{\sigma^{\mathrm{spill},\bk\bb}_n}{|\bb|}\right)^2
\end{align}
with $w_{\mathbf{e}_{\bb}}= w_\bb|\bb|^2$ becomes comparable among different systems.
By employing $\Tr\left\{\mathbb{1} - M^{\bk\bb}(M^{\bk\bb})^\dagger\right\}  \allowbreak= \allowbreak \sum_n(\sigma^{\mathrm{spill},\bk\bb}_n)^2$ and evaluating it for the infinitesimal case using Eq.~\eqref{eq:MkComplete}, we obtain Eq.~\eqref{eq:localSpillBZ}.

\section{Energy truncated mismatch measure} \label{app:Mcut}
\begin{figure}[pthb]
    \centering
    \includegraphics[width=0.99\columnwidth]{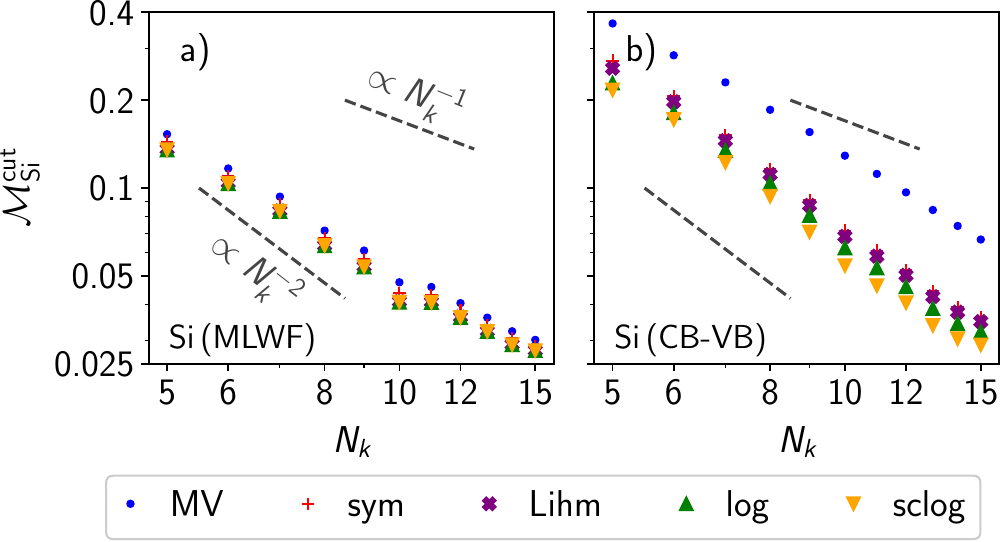}
    \caption{Energy truncated mismatch for a) MLWF and b) CB-VB parametrization of $\mathrm{Si}$ for the different Berry connection interpolation schemes as function of the ab-initio grid size $N_k$.}
    \label{fig:Mcut}
\end{figure}

The mismatch $\mathcal{M}$ for $\mathrm{Si}$ is dominated by contributions of the bands outside the frozen energy window used for the band disentanglement.
To show this, we define an truncated mismatch as 
\begin{align}
    M^{\mathrm{cut,S}} = \sqrt{\frac
                         {\int\limits_{\mathrm{BZ}} 
                         \sum\limits_{E_{i\bk},  E_{j\bk} \in [E^\mathrm{dis}_\mathrm{min}, E^\mathrm{dis}_\mathrm{max}]}
                         \left(\bv^{\mathrm{S,H},\bk}_{ij} - \bv^{\mathrm{abi,H},\bk}_{ij} \right)^2 \mathrm{d}\bk
                         }
                        { \int\limits_{\mathrm{BZ}} 
                         \sum\limits_{E_{i\bk},  E_{j\bk} \in [E^\mathrm{dis}_\mathrm{min}, E^\mathrm{dis}_\mathrm{max}]}
                         \left(\bv^{\mathrm{abi,H},\bk}_{ij}\right)^2 \mathrm{d}\bk}
                    }
,
\end{align}
where $E^\mathrm{dis}_\mathrm{min/max}$ represent the bounds of the frozen energy window used during disentanglement and $E_{i\bk}$ are the eigenenergies obtained from the Fourier-interpolated Hamiltonian.
Figure~\ref{fig:Mcut} shows the grid size dependent truncated energy mismatch, where the BZ integrals were evaluated on a $100\times 100\times 100$ $\bk$-grid.
The values of this energy truncated mismatch differ in two main aspects from the unrestricted ones defined in Eq.~\eqref{eq:mismatch}.
First, they are much smaller and second, they feature a quadratic scaling behavior with the same ordering of the schemes as obtained for $\mathrm{MoS}_2$, see Fig.~\ref{fig:pva}a and b.

\bibliography{refs}

\end{document}